\documentclass[preprint,prb]{revtex4}% Physical Review B
\usepackage{epsfig}
\usepackage{graphicx}% Include figure files
\usepackage{dcolumn}% Align table columns on decimal point
\usepackage{bm}% bold math

\begin{document}
\title{{\bf Multiscale Modeling of Materials - Concepts and Illustration}}
\author{Aditi Mallik, Keith Runge, James W. Dufty, and Hai-Ping Cheng}
\affiliation{Department of Physics, University of Florida,
Gainesville, FL 32611}
\date{\today }

%\draft
%\title{{\bf Multiscale Modelling of Materials - Concepts and Illustration }}
%\author{Aditi Mallik, Keith Runge, and James W. Dufty}
%\address{Department of Physics, University of Florida, Gainesville, Florida}

\begin{abstract}
The approximate representation of a quantum solid as an equivalent
composite semi-classical solid is considered for insulating
materials. The composite is comprised of point ions moving on a
potential energy surface. In the classical bulk domain this
potential energy is represented by pair potentials constructed to
give the same structure and elastic properties as the underlying
quantum solid. In a small local quantum domain the potential is
determined from a detailed quantum calculation of the electronic
structure. The formulation of this description as a sequence of
physical approximations is considered in some detail. The primary
new ingredients are 1) a determination of the pair potential from
quantum chemical data for equilibrium and strained structures, 2)
development of 'pseudo-atoms' for a realistic treatment of charge
densities where bonds have been broken to define the quantum
domain, and 3) inclusion of polarization effects on the quantum
domain due to its environment. This formal structure is
illustrated in detail for an $SiO_{2}$ nanorod. For each
configuration considered, the charge density of the entire solid
is calculated quantum mechanically to provide the reference by
which to judge the accuracy of the modeling. The construction of
the pair potential, the rod, the pseudo-atoms, and the multipoles
is discussed and tested in detail. It is then shown that the
quantum rod, the rod constructed from the classical pair
potentials, and the composite classical/quantum rod all have the
same equilibrium structure and response to elastic strain. In more
detail, the charge density and forces in the quantum subdomain are
accurately reproduced by the proposed modeling of the
environmental effects even for strains beyond the linear domain.
The accuracy of the modeling is shown to apply for two quite
different quantum chemical methods for the underlying quantum
mechanics: transfer Hamiltonian and density functional methods.
\end{abstract}

 \maketitle

\section{Introduction}

The field of multiscale modeling has opened a new era to
computational science for studying complex phenomena like
fracture, hydrolysis, enzymatic reactions, solute-solvent studies,
hydrogen embrittlement, and many other chemo-mechanical processes
in macroscopic samples. Often these phenomena require a very
accurate description at one scale, while at another scale a much
coarser method can be applied to get satisfactory results. In
fact, the coarser description for the bulk sample is required
because the more fundamental methods are computationally too
intensive to be applied to the entire system. These scales may be
length or time, or a combination of both. This scheme of combining
different models at different scales to achieve a balance of
accuracy, efficiency, and realistic description is known as
multiscale modeling. It is accomplished by applying the high
accuracy method only in a small domain where it is needed the most
and more approximate methods for the rest of the bulk where they
are appropriate. For example, in crack propagation one has to
apply detailed quantum mechanics at the tip of the crack. Here the
bonds are breaking between the atoms, causing a marked deformation
of the electron cloud and charge transfer between ions. But far
from the crack tip where the atoms are less deformed they can be
described by classical mechanics with appropriate potentials.
Thus, the problem arises of linking quite different descriptions
for the different length scales \cite{Broughton1999}.

There is a broad diversity of other examples requiring multiscale
modeling: chemo-mechanical polishing \cite{Singh2002}, tidal wave
prediction \cite{Clementi1988}, atmospheric sciences,
embrittlement of nuclear reactors \cite{Odette2001}, and many
biological systems \cite{Gogonea2002}. Two classes can be
distinguished \cite{Rudd2000}, ``serial multiscale modeling''
where the various scales are weakly coupled but the computation of
parameters at smaller scales is required for its use in more
phenomenological models at a larger scale, and ``concurrent
multiscale modeling'' where the various descriptions applied on
different scales should all be nested with proper boundary
conditions. The multiscale modeling discussed here is of the
``concurrent'' type, although there are components that can be
considered as the ``serial'' type (e.g., parameterization of the
underlying approximate quantum method used).

Three different length scale levels are distinguished, the
nanoscale where the details of electron structure and quantum
chemistry are important, the atomic or microscale where
appropriate classical pair potentials capture the structure
accurately, and the macroscale where a continuum mechanical
description applies. The present work addresses only one
particular subclass of multiscale modeling, known as hybrid
quantum mechanical (QM)/classical mechanical (CM) problems, and
their application to solid insulators - silica in particular.  The
further embedding of the atomic scale in a macroscale model is not
considered here. Early work on this QM/CM topic began with Warshel
and Levitt \cite{Warshel1976} and has accelerated over the past
decade\cite{Gao1996}.  Applications of QM/CM include biological
systems where this scheme is sometimes referred to as QM/MM
(molecular mechanics) modeling (enzymes, DNAs and proteins)
\cite{Monard1999,Amara1999,Cui2002,Titmuss2000,Murphy2002}%
, vibrational spectroscopy \cite{Chabana2001}, electronic
excitations \cite{Thompson1996,Gao1996,Vries1996}, hydrolysis of silica \cite{Jung2001,Du2003}%
, and solute-solvent problems
\cite{Eichinger1999,Guo1992,Gao1997}.

The challenge of the QM/CM simulation is to move from one length
scale to another, as smoothly as possible. A solid or large
cluster behaves as a single ''molecule'' so the partitioning
becomes more complicated due to covalent bonds that are cut
between the QM and CM regions. These dangling bonds give rise to
incorrect charge densities in the QM domain and other pathologies
\cite{Sauer2000,Ogata2004}. \  A multiscale modeling scheme is
proposed here that addresses this and other problems of the
quantum embedding to provide a means for faithful coupling between
the QM and CM regions. There are many tests for fidelity across
the QM/CM interface, such as bond angles, bond lengths, and proton
affinities. Instead, the criteria here are preservation of
accurate electronic charge densities and total forces in the QM
sub domain. These are the physically significant properties based
more closely in the theoretical structure being modeled, as
described in the next section. The approach includes the following
two components:

(i) Modeling environmental effects for the quantum domain in an
accurate and simple manner. Here, accurate means that the forces
and charge densities of the QM domain are reproduced to within a
few percent. This entails a chemically correct treatment of the
dangling bonds, and an account of longer range Coulomb influences
from the classical domain. Simplicity, means that the
approximations made are transparent and that the difficulty of the
quantum calculation is not increased. Here, the valency problem is
solved by pseudo-atoms and the longer range environmental effects
are included through low order polarization effects
\cite{Mallik2004}.

(ii) Developing a new classical potential for the classical region
tailored to the properties of interest. A new potential for the
entire CM region is designed, based on selected force data
calculated from the quantum method used in the QM domain. This
data is generated for both equilibrium and near equilibrium
states. The test for a suitable potential, in addition to
producing the right structure, is the accurate reproduction of the
chosen property to be studied, for near equilibrium states
(typically some linear response characteristic)
\cite{Mallik20042}. The use of this new potential prevents any
mismatch of that response property across the QM/CM boundary.
Hence different problems (fracture or corrosion) might require the
construction of different potentials for the same system.

The criterion for the success of (i) and (ii) is that calculations
of the desired properties  should be indistinguishable for the
quantum system, purely classical system, and the composite QM/CM
system for near equilibrium states. Only then are the desired
applications to far from equilibrium states reliable.

\begin{figure}[ht]
\includegraphics[width=4cm]{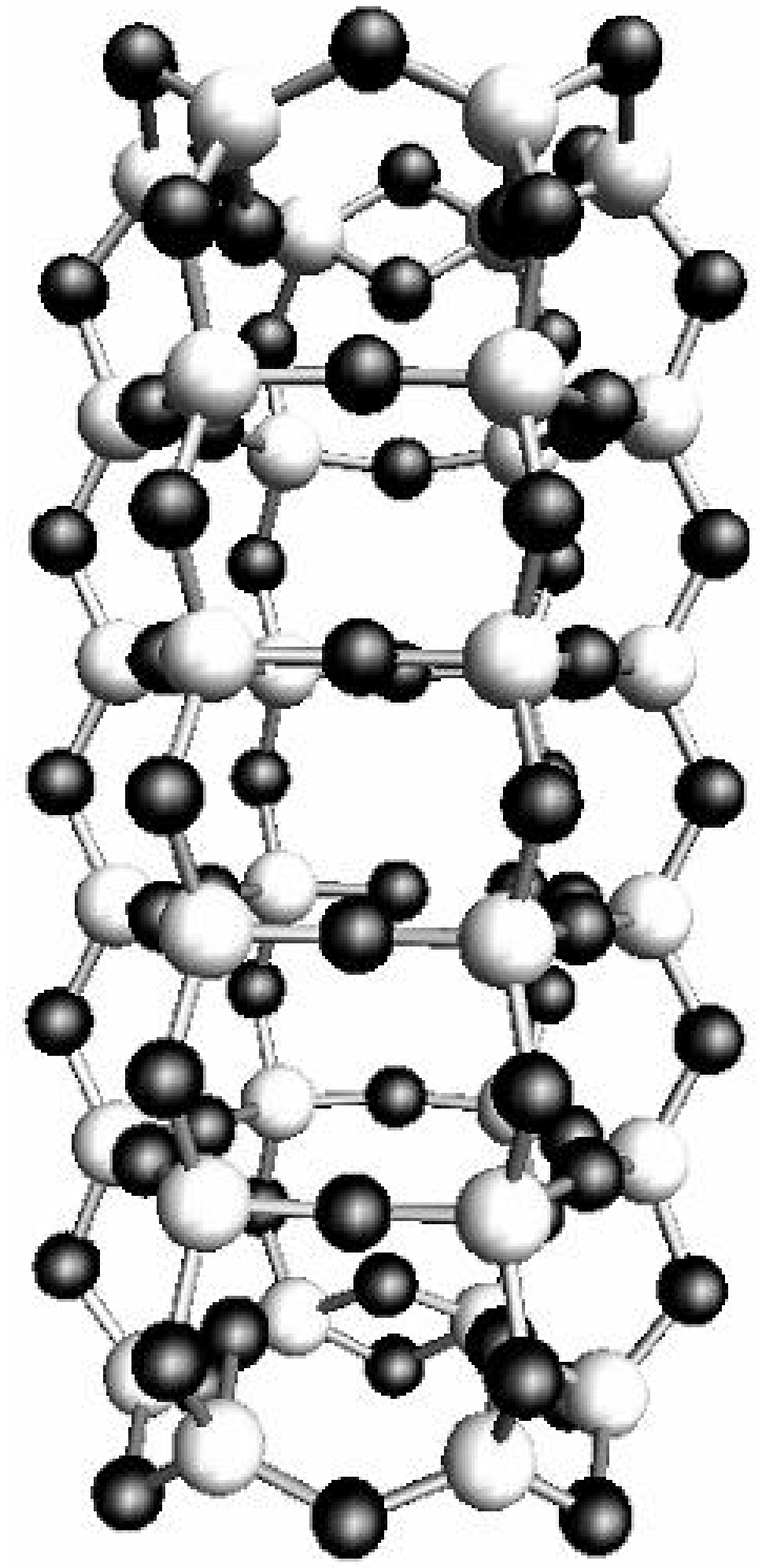} \hspace{2.0cm}
\includegraphics[width=9cm]{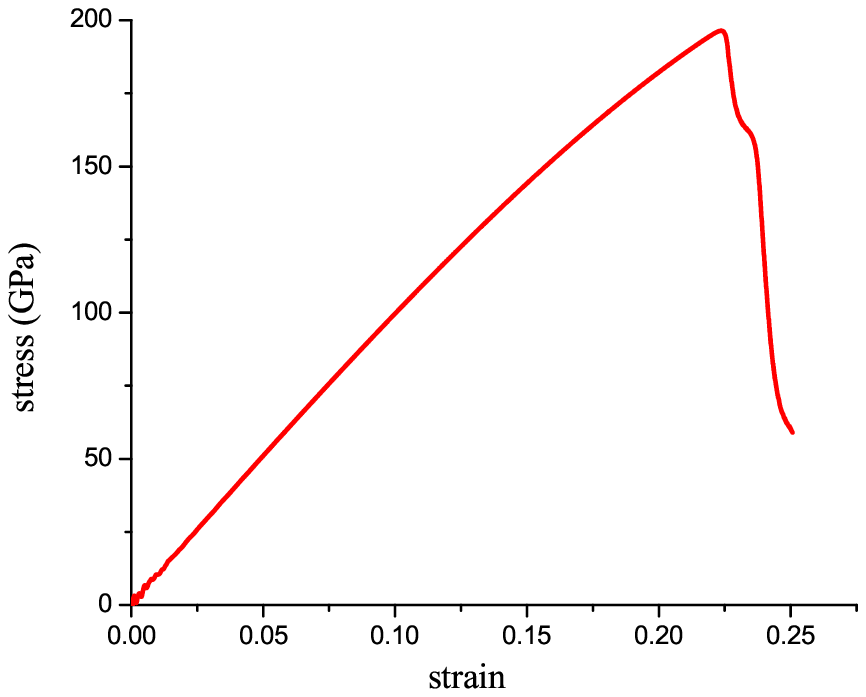}
\caption{( color online) Structure of the $SiO_{2}$ nanorod (left), and quantum
stress - strain curve showing bulk solid properties (right)}
\label{fig1}
\end{figure}

In the next four sections, the theoretical basis for this modeling
scheme is described in some detail. The object is to provide a
means to assess the approximations made more directly, and also to
provide the basis for their extension to other systems. The
complete formalism is then tested for a model system - a silica
nanorod \cite{Zhu2003} (Figure \ref{fig1}) - in section VI. This
silica nanorod containing 108 nuclei has been chosen since a
quantum treatment of the whole sample is possible to assess
quantitatively all aspects of the proposed scheme. Still, the
system is big enough to yield bulk properties like stress-strain
response (also shown in Figure \ref{fig1}). The nanorod has the
proper stoichiometric ratio of silicon to oxygen observed in real
silica (1:2) and is considered a viable model for studying silica.
One of the rings near the center of the rod is chosen to be the QM
domain and rest of the rod is treated classically (see Figure
\ref{fig6} below).

The system is studied under uniaxial strain. The quantum
mechanical method used is the Transfer Hamiltonian (TH) method
\cite{Taylor2003} using an NDDO type theory parameterized by
coupled cluster data (described more completely in section VI).
The construction of the classical pair potential is described and
shown to yield the same Young's constant as obtained from the QM
calculation, to within a few percent. Next, the training of the
pseudo atoms and validity of polarization representation for the
environment of the QM system is tested. The resulting force and
charge densities for the QM domain are accurately reproduced to
within a few percent. Finally, the composite QM/CM rod is shown to
have the same structure and elastic properties as both the QM and
CM rods for near equilibrium states, as required. The entire
analysis is repeated using density functional theory (DFT) as the
underlying quantum theory, instead of the TH quantum mechanics.
The same level of accuracy for the modeling is found to hold in
this case as well.

\section{The Idealized Quantum Solid}

\label{sec2}

In this section the problem of multiscale modeling is introduced by first
stating clearly what is the system being modeled. While much of this
introductory matter is familiar, it provides the context for assessing \emph{%
all} approximations involved in the final model material. First,
the quantum description in terms of ions and electrons is given
and the limitations for practical implementation are noted. Next,
reduced self-consistent descriptions are given for the ions and
electrons, each coupled to the other through their average charge
density. In this form the classical limit for the ions can be
taken, allowing the use of MD simulation methods for their
dynamics. The equation for the electrons is simplified in a
different direction, exploiting the very different time scales for
electron and nuclear motion (the Born-Oppenheimer limit). This
final description constitutes the idealized quantum solid for
which the subsequent multiscale modeling scheme is proposed.

At the fundamental level a simple atomic or molecular solid can be described
in terms of $N_{i}$ ''ions'' with charge number $Z_{i}$ for the
corresponding atoms of species $i$ and a set of $N_{e}$ electrons, with
overall charge neutrality ($\sum_{i}N_{i}Z_{i}=N_{e}$). To introduce the
various levels of description it is useful to start with the density
operator $D$ for the system as a whole. Properties of interest $A$ are given
by the expectation value%
\begin{equation}
\left\langle A\right\rangle =Tr_{e,i}DA,  \label{2.1}
\end{equation}%
where the trace is taken over all electron and ion degrees of freedom. The
density operator obeys the Liouville - von Neumann equation%
\begin{equation}
\partial _{t}D+\frac{i}{\text{%
%TCIMACRO{\U{127}}%
%BeginExpansion
h{\hskip-.2em}\llap{\protect\rule[1.1ex]{.325em}{.1ex}}{\hskip.2em}%
%EndExpansion
}}[H,D]=0.  \label{2.2}
\end{equation}%
The Hamiltonian operator $H$ is comprised of the Hamiltonians $H_{i}$ and $%
H_{e}$ for the isolated systems of ions and electrons, respectively, and
their interaction $U_{ie}$
\begin{equation}
H=H_{i}+H_{e}+U_{ie},  \label{2.3}
\end{equation}%
\begin{equation}
U_{ie}=\int d\mathbf{r}d\mathbf{r}^{\prime }\frac{\rho _{i}\left( \mathbf{r}%
\right) \rho _{e}\left( \mathbf{r}^{\prime }\right) }{\left| \mathbf{r}-%
\mathbf{r}^{\prime }\right| }.  \label{2.4}
\end{equation}%
Here $\rho _{i}\left( \mathbf{r}\right) $ and $\rho _{e}\left( \mathbf{r}%
\right) $ are the ion and electron charge density operators. This
is the most fundamental level for a quantum description of the
system. For a pure state, $D$ is a projection onto a state $\Psi $
which is determined from the corresponding Schroedinger equation.
For small systems, this can be solved numerically by quantum Monte
Carlo methods. At this level, the approximations involved (e.g.,
electron nodal location) can be considered mild and under control.
However, quantum Monte Carlo methods are restricted to relatively
small systems and the above fundamental description is not a
practical method for application to larger systems, particularly
if repeated calculation is required to follow the dynamics.

In many cases the properties of interest are functions only of the
ion degrees of freedom (e.g. structure), $A\rightarrow A_{i}$, or
they are purely electronic (e.g., optical), $A\rightarrow A_{e}$.
Then a description is possible in terms the reduced density
operators for the ions and for the electrons, $D_{i}$ and $D_{e}$
, resulting from appropriate partial traces
over all the electrons or ions, respectively%

\begin{equation}
\left\langle A_{i}\right\rangle
=Tr_{i}D_{i}A_{i},\hspace{0.3in}\left\langle A_{e}\right\rangle
=Tr_{e}D_{e}A_{e},  \label{2.5}
\end{equation}%
\begin{equation}
D_{i}\equiv Tr_{e}D,\hspace{0.3in}D_{e}\equiv Tr_{i}D.
\label{2.5a}
\end{equation}%
Their equations follow directly from (\ref{2.2})%
\begin{equation}
\partial _{t}D_{i}+\frac{i}{\text{%
%TCIMACRO{\U{127}}%
%BeginExpansion
h{\hskip-.2em}\llap{\protect\rule[1.1ex]{.325em}{.1ex}}{\hskip.2em}%
%EndExpansion
}}[H_{i},D_{i}]+\frac{i}{\text{%
%TCIMACRO{\U{127}}%
%BeginExpansion
h{\hskip-.2em}\llap{\protect\rule[1.1ex]{.325em}{.1ex}}{\hskip.2em}%
%EndExpansion
}}\left( U_{i}D_{i}-D_{i}U_{i}^{\dagger }\right) =0,  \label{2.6}
\end{equation}%
\begin{equation}
\partial _{t}D_{e}+\frac{i}{\text{%
%TCIMACRO{\U{127}}%
%BeginExpansion
h{\hskip-.2em}\llap{\protect\rule[1.1ex]{.325em}{.1ex}}{\hskip.2em}%
%EndExpansion
}}[H_{e},D_{e}]+\frac{i}{\text{%
%TCIMACRO{\U{127}}%
%BeginExpansion
h{\hskip-.2em}\llap{\protect\rule[1.1ex]{.325em}{.1ex}}{\hskip.2em}%
%EndExpansion
}}\left( U_{e}D_{e}-D_{e}U_{e}^{\dagger }\right) =0.  \label{2.9}
\end{equation}%
where the potential energy operators coupling the ion and
electronic
degrees of freedom are%
\begin{equation}
U_{i}=\int d\mathbf{r}d\mathbf{r}^{\prime }\frac{1}{\left| \mathbf{r}-%
\mathbf{r}^{\prime }\right| }\rho _{i}\left( \mathbf{r}\right) \widetilde{%
\rho }_{e}\left( \mathbf{r}^{\prime }\right) ,\hspace{0.3in}\widetilde{\rho }%
_{e}\left( \mathbf{r}\right) =\left( Tr_{e}\rho _{e}\left(
\mathbf{r}\right) D\right) \left( Tr_{e}D\right) ^{-1},
\label{2.7}
\end{equation}%
\begin{equation}
U_{e}=\int d\mathbf{r}d\mathbf{r}^{\prime }\frac{1}{\left| \mathbf{r}-%
\mathbf{r}^{\prime }\right| }\widetilde{\rho }_{i}\left(
\mathbf{r}\right)
\rho _{e}\left( \mathbf{r}^{\prime }\right) ,\hspace{0.3in}\widetilde{\rho }%
_{i}\left( \mathbf{r}\right) =\left( Tr_{i}\rho _{i}\left(
\mathbf{r}\right) D\right) \left( Tr_{i}D\right) ^{-1}
\label{2.10}
\end{equation}%
This is similar to the microscopic ion - electron coupling of
(\ref{2.4}) except that now the electron charge density operator
$\rho_{e} \left(\mathbf{r}\right) $ is replaced by its conditional
average, $\widetilde{\rho }_{e}\left( \mathbf{r}\right) $, in the
equation for $D_{i}$, and the ion charge density $\rho _{i}\left(
\mathbf{r}\right) $ is replaced by its conditional average,
$\widetilde{\rho }_{i}\left( \mathbf{r}\right) $ in the equation for $D_{e}$. The description (\ref{2.6}) and (\ref%
{2.9}) is still exact but formal since these average charge
densities are not determined by these equations alone. The
simplest realistic approximation (mean field) is to neglect the
direct correlations in the
charge densities $\widetilde{\rho }_{i}\left( \mathbf{r}\right) $ and $%
\widetilde{\rho }_{e}\left( \mathbf{r}\right) $, i.e. replace
$D\rightarrow D_{e}D_{i}$ in (\ref{2.6}) and (\ref{2.10}) to get
\begin{equation}
\widetilde{\rho }_{i}\left( \mathbf{r}\right) \rightarrow
Tr_{i}\rho
_{i}\left( \mathbf{r}\right) D_{i}\equiv \overline{\rho }_{i}\left( \mathbf{r%
}\right) ,\hspace{0.3in}\widetilde{\rho }_{e}\left(
\mathbf{r}\right)
\rightarrow Tr_{e}\rho _{e}\left( \mathbf{r}\right) D_{e}\equiv \overline{%
\rho }_{e}\left( \mathbf{r}\right) .  \label{2.15}
\end{equation}%
As a consequence, the potentials $U_{i}$ and $U_{e}$ become Hermitian and
the average charge densities are now self-consistently determined by (\ref%
{2.6}) and (\ref{2.9}). Self-consistency is required since $D_{e}$ and $%
D_{i} $ are functionals of the average charge densities $\overline{\rho }%
_{i}\left( \mathbf{r}\right) $ and $\overline{\rho }_{e}\left( \mathbf{r}%
\right) $, respectively.

The advantage of this reduced description is that the ions and
electrons are described by separate equations, reducing the degree
of difficulty of each. More importantly, this separation allows
the introduction of appropriate approximations for each. The large
differences in electron and ion masses imply corresponding
differences in time scales and thermal de Broglie wavelengths.
Consequently, the equation for the ions admits a classical limit
for the conditions of interest, while that for the electrons does
not. The classical limit of equation (\ref{2.6}) becomes
\begin{equation}
\partial _{t}D_{i}+\left\{ \left( H_{i}+U_{i}\left[ \overline{\rho }_{e}%
\right] \right) ,D_{i}\right\} =0,  \label{2.16a}
\end{equation}%
where $\left\{ ,\right\} $ now denotes a Poisson bracket operation, and $%
D_{i}$ is a function of the ion positions and momenta,
$D_{i}\rightarrow D_{i}\left( \{\mathbf{R}_{i\alpha
}\},\{\mathbf{P}_{i\alpha }\},t\right)$ (here $\alpha$ denotes a
specific ion). This equation now can be solved accurately and
efficiently by molecular dynamics simulation methods, even for
large systems. This is an essential step in almost all practical
descriptions of bulk materials whose importance cannot be
overstated.

Implementation of (\ref{2.16a}) still requires calculation of the
potential energy $U_{i}\left[ \overline{\rho }_{e}\right] $, which
in turn requires determination of the electronic charge density
from $D_{e}$. However, the general solution to (\ref{2.9}) is a
formidable problem: determination of the dynamics of $N_{e}$
electrons self-consistently in the presence of a changing ion
charge density. Two simplifications are made to bring this problem
under control. First, it is recognized that $\tau \left(
\partial _{t}D_{i}\right) D_{i}^{-1}<<1$, where $\tau$ is the
time scale for changes in the electron distribution function
$D_{e}.$ Second, it is assumed that only the lowest energy state
contributes to $D_{e}$ at any given time. The first approximation
constitutes the Born-Oppenheimer approximation while the second
approximation is appropriate for most structural studies, but must
be relaxed for optical studies involving electronic transitions.
In principle, (\ref{2.16a}) is solved in time steps $\Delta t$. At
each time step the electronic charge density is computed for the
ion configurations at that time step in order to recompute new
forces for the next time step. The electronic charge density
calculation is a ground state eigenvalue problem for the given ion
configuration.

To summarize, the final description of this idealized solid, it
consists of a set of point ions governed by the classical equation
(\ref{2.16a}) for their probability distribution $D_{i}\rightarrow
D_{i}\left( \{\mathbf{R}_{i\alpha
}\},\{\mathbf{P}_{i\alpha }\},t\right) $%
\begin{equation}
\partial _{t}D_{i}+\left\{ \left( H_{i}+U_{i}\left[ \overline{\rho }_{e}%
\right] \right) ,D_{i}\right\} =0,  \label{2.23}
\end{equation}%
\begin{equation}
U_{i}=\int d\mathbf{r}d\mathbf{r}^{\prime }\frac{1}{\left| \mathbf{r}-%
\mathbf{r}^{\prime }\right| }\rho _{i}\left( \mathbf{r}\right) \overline{%
\rho }_{e}\left( \mathbf{r}^{\prime },t\right)
,\hspace{0.3in}\overline{\rho }_{e}\left( \mathbf{r},t\right)
=Tr_{e}\rho _{e}\left( \mathbf{r}^{\prime }\right) D_{e}\left[
\overline{\rho }_{i}\right] \label{2.24}
\end{equation}%
The average electron density $\overline{\rho }_{e}\left( \mathbf{r}\right) $
is determined from the ground state solution to (\ref{2.9})
\begin{equation}
\frac{i}{\text{%
%TCIMACRO{\U{127}}%
%BeginExpansion
h{\hskip-.2em}\llap{\protect\rule[1.1ex]{.325em}{.1ex}}{\hskip.2em}%
%EndExpansion
}}[H_{e}+U_{e},D_{e}]=0,  \label{2.25a}
\end{equation}%
\begin{equation}
U_{e}=\int d\mathbf{r}d\mathbf{r}^{\prime }\frac{1}{\left| \mathbf{r}-%
\mathbf{r}^{\prime }\right| }\overline{\rho }_{i}\left( \mathbf{r},t\right)
\rho _{e}\left( \mathbf{r}^{\prime }\right) ,\hspace{0.3in}\overline{\rho }%
_{i}\left( \mathbf{r},t\right) =Tr_{e}\rho _{i}\left( \mathbf{r}\right) D_{i}%
\left[ \overline{\rho }_{e}\right] \label{2.26}
\end{equation}%
The analysis proceeds stepwise. The classical equations (\ref%
{2.23}) are solved analytically in discrete time steps for the
atomic coordinates and momenta. At each step the electron problem
(\ref{2.25a}) is solved for the electron ground state using the
ion configuration at the previous time step. From this ground
state the electron charge density $\rho _{e}\left(
\mathbf{r}\right) $ is determined. This gives the potential energy
function $U_{i}$ and consequently the forces required to change
the ion positions and momenta at the next time step. The process
is repeated with a new electron charge density calculated at each
time step using the new ionic configurations. All electron
correlations are accounted for quantum mechanically in the
eigenvalue problem; all atomic correlations are determined
classically through direct solution of Newton's equations. All
structural properties of interest can be calculated since the
phase points for the ions are known at all times. The dynamics of
the electrons is only coarse-grained as they are "slaved" to the
time dependence of the ions.

The above defines a ''quantum molecular dynamics'' representation
of the idealized quantum solid. The semi-classical approximation
for the ions, and ground state Born-Oppenhiemer approximation for
the electrons are relatively weak under most conditions of
interest. The resulting description allows an accurate classical
treatment of the ionic structure while retaining relevant quantum
chemistry for the interatomic forces due to electrons. If it could
be implemented in practice for bulk systems of interest over
reasonable time intervals there would be little need for
multiscale modeling. With MD simulation the solution to
(\ref{2.23}) once $U_{i}$ has been provided is straightforward. So
the problem has been reduced to a determination of the electron
charge density. Unfortunately, the solution to (\ref{2.25a}) using
realistic quantum chemical methods for even a few hundred ions at
each time step becomes prohibitively time intensive.

\section{The Formal Partition and Composite Solid}

\subsection{The Representative Classical Solid}

In many cases of interest (e.g., equilibrium structure,
thermodynamics) the computational limitations of quantum chemical
methods can be avoided through a purely classical representation
of the solid, avoiding the intensive electron charge density
calculation at each time step. This entails an idealization that
has many variants. It consists of the
representation of the true potential energy function $U_{i}\left( \{\mathbf{R%
}_{i\alpha }\}\right) $ in (\ref{2.23}) by a suitably
\emph{chosen} function $U_{c}\left( \{\mathbf{R}_{i\alpha
}\}\right) $. Consequently, its form does not need to be computed
at each time step and the speed and efficiency of classical
molecular dynamics is not compromised.

The problem with this approach lies in the choice for $U_{c}\left( \{\mathbf{%
R}_{i\alpha }\}\right) $. In principle, an exact mapping for some
fundamental property such as the free energy $F$ can be imposed%
\begin{equation}
F[U_{i}]=F_{c}[U_{c}].  \label{3.1}
\end{equation}%
where $F_{c}$ is the corresponding classical functional.
Generally, such exact methods can be inverted only perturbatively
and lead to a sequence of effective many ion interactions
involving increasingly more particles. A more practical method is
to assume pairwise additivity for effective point
''atoms''%
\begin{equation}
U_{c}\left( \{\mathbf{R}_{i\alpha }\}\right) \rightarrow \frac{1}{2}%
\sum_{k,j}\sum_{\alpha }^{N_{k}}\sum_{\beta }^{N_{j}}V_{kj}\left(
\left| \mathbf{R}_{k\alpha }-\mathbf{R}_{j\beta }\right| \right) ,
\label{3.2}
\end{equation}%
where $k,i$ label the species (ion or electron). The exact determination of the pair
potentials $V_{kj}\left( \left| \mathbf{R%
}_{k\alpha }-\mathbf{R}_{j\beta }\right| \right) $ is now much
more restrictive as not all properties of interest have such a
representation.
Nevertheless pair properties such as the radial distribution functions $%
g_{kj}\left( \left| \mathbf{r}\right| \right) $ might be used
to determine the pair potentials. As the $g_{kj}\left( \left| \mathbf{%
r}\right| \right) $ are unknown and difficult to calculate, the
inversion is again difficult and not practical. Instead,
experimental data is often used to fit a parameterized functional
form chosen for the pair potentials. At this stage control over
the approximation is lost and the method becomes phenomenological.

This phenomenological approach has been and remains a valuable
tool of materials sciences. However, in the context of multiscale
modeling it must be reconsidered carefully. First, it is
recognized that there are local domains far from equilibrium where
a purely classical potential cannot apply because of the inherent
quantum chemistry active there (e.g., charge transfer and
exchange). Conversely, there are large complementary domains in
near equilibrium states where representation by an appropriate
classical potential is possible. Multiscale modeling constructs
distinct classical and quantum models of these subsystems and then
requires fidelity at their interface. This is a severe test of the
modeling assumptions in each subsystem. The approach proposed here
confronts this issue directly in the construction of appropriate
pair potentials for the problem considered.

\subsection{The Composite Quantum/Classical Solid}

It is presumed that there is some method for identifying domains
within the solid for which quantum chemical effects should be
treated in detail. The quantum solid is then represented as a
composite of two domains, the larger bulk in which a classical
representation is used and a smaller ''reactive'' domain in which
the original quantum description is retained. The objective of the
modeling described here is therefore to construct the potential
function $U_{i}$ such that it gives an accurate description in
both the reactive and non-reactive domains. This has two
components, the determination of a pair potential for the forces
on ions in the classical domain, and an accurate calculation of
the charge density for the forces on ions in the quantum domain.

The classical and quantum domains are defined by labelling all
ions as either classical or quantum, and associating spatial
domains with the
coordinates of each, denoted $\{\mathbf{R}_{ci\alpha }\}$ and $\{\mathbf{R}%
_{qi\alpha }\}$ respectively. The quantum domains are assumed to
be small, to allow practical calculation of the electronic
structure. In principle there could be several disconnected
quantum domains, but to simplify the discussion we consider only
one. It is assumed that initially the two sets are contiguous with
a smooth interface and that diffusion or migration between them is
not significant over the times of interest. In addition to the
ions in the quantum domain, there are $m$ $\ $electrons where $m$
is determined by a condition on the charge of the quantum domain,
here taken to be neutral. The
total average electron charge density is then decomposed as%
\begin{equation}
\overline{\rho }_{e}\left( \mathbf{r},t\right) =\overline{\rho
}_{e}\left( \mathbf{r},t\right) \left( \chi _{\mathcal{Q}}\left(
\mathbf{r}\right) +\chi
_{\mathcal{C}}\left( \mathbf{r}\right) \right) \equiv \overline{\rho }%
_{eq}\left( \mathbf{r},t\right) +\overline{\rho }_{ec}\left( \mathbf{r}%
,t\right) ,  \label{3.3}
\end{equation}%
where $\chi _{\mathcal{Q}}\left( \mathbf{r}\right) $ and $\chi _{\mathcal{C}%
}\left( \mathbf{r}\right) $ are characteristic functions for
$\mathcal{Q}$ and $\mathcal{C}$. The boundaries of the quantum
spatial domain $\mathcal{Q}$ are constrained by the choice of ions
and total electron charge
\begin{equation}
me=\int_{\mathcal{Q}}d\mathbf{r}\overline{\rho }_{eq}\left( \mathbf{r}%
,t\right) ,  \label{3.4}
\end{equation}%
where $\mathcal{Q}$ encloses all $\{\mathbf{R}_{qi\alpha }\}$
(this leaves some flexibility to define smooth interfaces). The
complement of this is the classical domain $\mathcal{C}$. This
gives a corresponding decomposition of the total ion potential
energy function for the ions $V_{i}$ (ion - ion Coulomb
interactions plus $U_{i}$ of (\ref{2.24})) so that (\ref{2.23})
becomes
\begin{equation}
\partial _{t}D_{i}+\left\{ \left( K_{i}+V_{ic}+V_{iq}\right) ,D_{i}\right\}
=0,  \label{3.5}
\end{equation}%
where $K_{i}$ is the ion kinetic energy and
\begin{equation}
V_{ic}=\frac{1}{2}\int_{\mathcal{C}}d\mathbf{r}\rho _{i}\left( \mathbf{r}%
\right) \left( \int_{\mathcal{C}}d\mathbf{r}^{\prime
}\frac{1}{\left|
\mathbf{r}-\mathbf{r}^{\prime }\right| }\left( \rho _{i}\left( \mathbf{r}%
^{\prime }\right) -\delta \left( \mathbf{r-r}^{\prime }\right) +\overline{%
\rho }_{ec}\left( \mathbf{r}^{\prime },t\right) \right) +\int_{\mathcal{Q}}d%
\mathbf{r}^{\prime }\frac{1}{\left| \mathbf{r}-\mathbf{r}^{\prime }\right| }%
\left( \rho _{i}\left( \mathbf{r}^{\prime }\right) +\overline{\rho }%
_{eq}\left( \mathbf{r}^{\prime },t\right) \right) \right) ,
\label{3.6}
\end{equation}%
and%
\begin{equation}
V_{iq}=\frac{1}{2}\int_{\mathcal{Q}}d\mathbf{r}\rho _{i}\left( \mathbf{r}%
\right) \left( \int_{\mathcal{Q}}d\mathbf{r}^{\prime
}\frac{1}{\left|
\mathbf{r}-\mathbf{r}^{\prime }\right| }\left( \rho _{i}\left( \mathbf{r}%
^{\prime }\right) -\delta \left( \mathbf{r-r}^{\prime }\right) +\overline{%
\rho }_{eq}\left( \mathbf{r},t\right) \right) +\int_{\mathcal{C}}d\mathbf{r}%
^{\prime }\frac{1}{\left| \mathbf{r}-\mathbf{r}^{\prime }\right|
}\left( \rho _{i}\left( \mathbf{r}^{\prime }\right)
+\overline{\rho }_{ec}\left( \mathbf{r}^{\prime },t\right) \right)
\right) .  \label{3.7}
\end{equation}%
The first terms of the integrands on the right sides of (\ref{3.6}) and (\ref%
{3.7}) represent the interactions of the ions with a given
subsystem in the presence of the average electronic charge density
of that subsystem. The second terms represent the interactions of
those ions with their complementary subsystem. Half of the ion -
ion potential energy between the two subsystems has been
associated with each potential in this decomposition so that the
total force acting on the quantum domain by the classical domain
is equal and opposite to that on the classical domain due to the
quantum domain.

The potential $V_{ic}$ is due to ions. By definition these ions
are in near equilibrium states and therefore this part of the
potential  should be represented well by classical pair potentials
of the form (\ref{3.2}), with appropriately chosen parameters as
discussed below. The potential $V_{iq}$ is due to ions in the
quantum domain. As this is the domain that can be far from
equilibrium the average electronic charge densities must be
calculated in detail from the quantum description (\ref{2.25a}).
There are two parts to this charge density affecting the ions in
the quantum domain, that due to
the $m$ electrons of the quantum domain $\overline{\rho }_{eq}\left( \mathbf{%
r},t\right) $, and that due to the surrounding classical domain
$\rho
_{i}\left( \mathbf{r}\right) +\overline{\rho }_{ec}\left( \mathbf{r}%
,t\right) $. The equation governing the  $m$ electrons of the
quantum domain is coupled to this same classical domain average
charge density. The scheme
proposed here consists of modeling this charge density $\overline{\rho }%
_{ec}\left( \mathbf{r},t\right) $ as an accurate and practical
representation for the environment of the quantum domain, allowing
calculation of $\overline{\rho }_{eq}\left( \mathbf{r},t\right) $
and therefore determining both  $V_{iq}$ and $V_{ic}$.

In summary, the multiscale model is obtained by replacing the ion
- ion contribution in the classical domain (the first term on the
right side of (\ref{3.6})) by suitably parameterized pair
potentials, and constructing an accurate approximate calculation
of the electron density in the quantum domain. The latter entails
a representation of the effects of the electron density in the
classical domain on charges in the quantum domain. In this way the
potentials of (\ref{3.6}) and (\ref{3.7}) are entirely determined.
The details of this construction are addressed in the following
sections.

\section{Description of the Classical Subsystem}

\subsection{Construction of the pair potentials}

The proposed method for constructing a classical pair potential
for use in multiscale modeling has several components: 1) it
should be constructed for accuracy of the specific properties to
be studied, 2) it should be ''trained'' on quantum data generated
in the same way as for the quantum domain at its interface, 3) it
should predict accurate equilibrium structure, but include
training on appropriate near equilibrium states as well, and 4)
simplicity of form for parameterization and implementation in MD
codes should be maintained.

The steps in constructing a potential are the following. First the
specific quantum method to be used in the multiscale modeling is
identified (below, transfer Hamiltonian or density functional) and
applied to a large cluster or representative sample of the solid
to be modeled. The forces on atoms for both equilibrium and near
equilibrium states are then calculated quantum mechanically. Next,
a simple functional form for the pair potentials with the correct
physical shape (e.g., one of the existing phenomenological forms)
is chosen and the parameters controlling that shape selected for
optimization. The forces on the ions are calculated for these
chosen potentials at the configurations used in the quantum force
calculations, and compared with those quantum forces. The
parameters are adjusted for a best fit to the quantum force data.
When a good fit has been obtained, the potential energy at
equilibrium is tested for stability using a gradient algorithm to
establish that a local minimum of the potential has been obtained.
Finally, the property of interest (e.g., some linear response) is
tested by comparing its calculation using MD simulation with the
fitted potentials and that with the quantum forces. If necessary,
the fitting procedure can be repeated with differing weights for
the quantum force data in equilibrium and near equilibrium states,
or other additional input from the quantum calculations.

A more detailed discussion of the flexibility in this procedure is
described for the example of the next section. It entails some art
as well as science in choosing the optimization methods and
applying them. In that example, it was found that some repeated
combination of genetic algorithms and scaling provided the
accuracy required. The primary result of this approach is a
classical pair potential which gives the properties of interest by
design, that match those being calculated quantum mechanically
across the interface - for near equilibrium states.

\subsection{Effects of the environment}

The above construction describes the modeling of the first term on
the right side of (\ref{3.6}) where both the ions and the average
electron charge density refer to the classical domain, assumed in
a near equilibrium state. The second term involves a coupling to
ions and average charge density (now the ground state charge
density) in the quantum domain $\overline{\rho }_{eq}\left(
\mathbf{r},t\right) $. This charge density is computed for the
forces in the quantum domain (next section) and hence the second
term of (\ref{3.6}) is known as well. Thus, the potential energy
for the ions of the classical domain is determined from
synthesized pair potentials among the ions and electrons of the
classical domain, plus an interaction with the electrostatic
potential of the ions and average electron charge density of the
quantum domain. However, in the examples discussed below the
coupling of the classical domain ions to the quantum domain is
simplified by using the same pair potentials as for ions within
the classical domain. This is expected to be quite accurate if the
quantum domain charge density is not distorted very much from the
near equilibrium states, since the potentials are trained to be
equivalent to the charge density in the near equilibrium domain.

\section{Description of the Quantum Subsystem}

The quantum domain is a subsystem of $m$ electrons localized about
the designated ions defining that domain. Consider the reduced
density operator for $m$ electrons defined by
\begin{equation}
D_{e}^{(m)}=Tr_{e}^{N_{e}-m}D_{e}  \label{5.1}
\end{equation}%
More specifically, this partial trace is defined in coordinate
representation by%
\begin{equation}
\left\langle \mathbf{r}_{1},..,\mathbf{r}_{m}\right|
D_{e}^{(m)}\left|
\mathbf{r}_{1}^{\prime },..,\mathbf{r}_{m}^{\prime }\right\rangle =\int d%
\mathbf{r}_{m+1}..d\mathbf{r}_{N_{e}}\left\langle \mathbf{r}_{1},..,\mathbf{r%
}_{m},\mathbf{r}_{m},..\mathbf{r}_{N_{e}}\right| D_{e}\left| \mathbf{r}%
_{1}^{\prime },..,\mathbf{r}_{m}^{\prime },\mathbf{r}_{m},..\mathbf{r}%
_{N_{e}}\right\rangle  \label{5.2}
\end{equation}%
Clearly the full exchange symmetry among all $N_{e}$ electrons is
preserved. However, this reduced density operator is not specific
to the quantum domain
defined above only. For example, the diagonal elements $\left\langle \mathbf{%
r}_{1},..,\mathbf{r}_{m}\right| D_{q}\left| \mathbf{r}_{1},..,\mathbf{r}%
_{m}\right\rangle $ give the probability density to find $m$
electrons at the specified positions, and the latter can be chosen
anywhere inside the
system. Thus, only when the positions are restricted to the quantum domain $%
\mathcal{Q}$ does this reduced density operator represent the
electrons of
that domain. Similarly, if $\rho _{e}^{\left( m\right) }\left( \mathbf{r}%
\right) $ is the charge density operator for $m$ electrons its
average is
\begin{equation}
\overline{\rho }_{e}^{(m)}\left( \mathbf{r}\right) =Tr_{e}\rho
_{e}^{\left( m\right) }\left( \mathbf{r}\right)
D_{e}=Tr_{e}^{m}\rho _{e}^{\left( m\right) }\left(
\mathbf{r}\right) D_{e}^{(m)}  \label{5.3}
\end{equation}%
where the trace in the second equality is over $m$ degrees of
freedom. This average charge density represents the average
contribution of $m$ electrons
at any point $\mathbf{r}$. Both $\left\langle \mathbf{r}_{1},..,\mathbf{r}%
_{m}\right| D_{q}\left| \mathbf{r}_{1},..,\mathbf{r}_{m}\right\rangle $ and $%
\overline{\rho }_{e}^{(m)}\left( \mathbf{r}\right) $ change with
$\mathbf{r} $ since there is an absolute reference background set
by the functional dependence on the ion charge density.
Consequently, in all of the following
discussion of this section $\left\langle \mathbf{r}_{1},..,\mathbf{r}%
_{m}\right| D_{q}\left| \mathbf{r}_{1},..,\mathbf{r}_{m}\right\rangle $ and $%
\overline{\rho }_{e}^{(m)}\left( \mathbf{r}\right) $ are
considered only for positions within the chosen quantum domain.
Accordingly, this coordinate representation $\left\{ \left|
\mathbf{r}_{1},..,\mathbf{r}_{m}\right\rangle \right\} $ defines
an $m$ electron Hilbert space of functions defined over the
quantum domain $\mathcal{Q}$. In this context $D_{e}^{(m)}$
becomes the reduced density operator for the quantum domain and
$\rho _{e}^{\left(
m\right) }\left( \mathbf{r}\right) $ its charge density operator%
\begin{equation}
D_{e}^{(m)}\rightarrow D_{q},\hspace{0.3in}\rho _{e}^{\left(
m\right) }\left( \mathbf{r}\right) \rightarrow \rho _{eq}\left(
\mathbf{r}\right) . \label{5.4}
\end{equation}

The equation determining the reduced density operator for the
quantum
subsystem follows directly from this definition and Eq. (\ref{2.25a}) for $%
D_{e}$%
\begin{equation}
\frac{i}{\text{%
%TCIMACRO{\U{127}}%
%BeginExpansion
h{\hskip-.2em}\llap{\protect\rule[1.1ex]{.325em}{.1ex}}{\hskip.2em}%
%EndExpansion
}}\left[ \left( K_{e}+V_{eq}\right) ,D_{q}\right] =0.  \label{5.5}
\end{equation}%
where $K_{e}$ is the kinetic energy for the $m$ electrons and
\begin{equation}
V_{eq}=\frac{1}{2}\int_{\mathcal{Q}}d\mathbf{r}\rho _{eq}\left( \mathbf{r}%
\right) \left( \int_{\mathcal{Q}}d\mathbf{r}^{\prime
}\frac{1}{\left|
\mathbf{r}-\mathbf{r}^{\prime }\right| }\left( \rho _{eq}\left( \mathbf{r}%
^{\prime }\right) -\delta \left( \mathbf{r}-\mathbf{r}^{\prime }\right) +%
\overline{\rho }_{i}\left( \mathbf{r}^{\prime }\right) \right) +\int_{%
\mathcal{C}}d\mathbf{r}^{\prime }\frac{1}{\left| \mathbf{r}-\mathbf{r}%
^{\prime }\right| }\left( \overline{\rho }_{ec}\left(
\mathbf{r}^{\prime }\right) +\overline{\rho }_{i}\left(
\mathbf{r}^{\prime }\right) \right) \right) .  \label{5.6}
\end{equation}%
(The contribution from $\overline{\rho }_{ec}\left(
\mathbf{r}^{\prime }\right) $ in the second term actually should
be a conditional average; the same mean field approximation as in
(\ref{2.15}) has been introduced here for consistency). The first
term on the left side of (\ref{5.6}) describes the isolated
quantum subsystem, composed of the Coulomb interactions among the
$m$ electrons \ and their coupling to the average charge density
of the ions in the quantum domain. The second term is the
interaction of these electrons with their environment, the total
average charge density of the classical domain.

There are two distinct types of contributions from this charge
density of the classical domain. The first is associated with a
subset of ions at the border of the QM/CM domains which describe
chemical bonds in the full quantum solid. These ions locate
regions where there is a highly localized electron charge density
shared with the quantum domain, including both strong correlation
and exchange effects, localized and non-uniform. The second type
of contribution is associated with the remaining border ions and
those charges more distant. In this second case the quantum
correlation and exchange effects of the first type are much
weaker, and the dominant effect is that of a polarized charge
density due to the ions and electrons.

These two types of effects of the environment can be identified by
separating the total charge density for the classical domain in
(\ref{5.6}) into domains centered on the border ions responsible
for bonding, and their
complement%
\begin{equation}
\overline{\rho }_{c}\left( \mathbf{r}\right) \equiv \overline{\rho }%
_{ec}\left( \mathbf{r}\right) +\overline{\rho }_{i}\left(
\mathbf{r}\right)
=\sum_{\alpha \subset \mathcal{V}}\chi \left( \left| \mathbf{r}-\mathbf{R}%
_{\alpha }\right| \right) \overline{\rho }_{c}\left(
\mathbf{r}\right) +\Delta \overline{\rho }_{c}\left(
\mathbf{r}\right)  \label{5.7}
\end{equation}%
where it is understood that $\mathbf{r}$ is in the classical
domain. The set of border ions for which a bond has been broken in
identifying the quantum domain is denoted by $\mathcal{V}$.
Also, $\chi \left( \left| \mathbf{r}-%
\mathbf{R}_{\alpha }\right| \right) $ is a characteristic function
specifying a domain centered on $\mathbf{R}_{\alpha }$ such that
it does not overlap neighboring ions. Its size is taken large
enough to incorporate the bound electrons forming the ''atom'' for
this ion in the quantum solid. \ The second term $\Delta
\overline{\rho }_{c}\left( \mathbf{r}\right) $ is the charge
density for all remaining ions and electrons of the classical
environment.

\subsection{Coulomb effects of environment}

By its definition, the averaged charge density $\Delta \overline{\rho }%
_{c}\left( \mathbf{r}\right) $ does not include the contributions
to chemical bonding with the quantum domain required for its
valency. Hence the electrostatic potential associated with it can
be expected to have a regular multipole expansion
\begin{equation}
\int_{\mathcal{Q}}d\mathbf{r}^{\prime }\frac{1}{\left| \mathbf{r}-\mathbf{r}%
^{\prime }\right| }\Delta \overline{\rho }_{c}\left(
\mathbf{r}\right) \rightarrow \frac{\widehat{\mathbf{r}}\cdot
\mathbf{d}}{r^{2}}+.. \label{5.8}
\end{equation}%
The leading monopole term is zero due to charge neutrality of the
classical
subsystem, and the dipole moment $\mathbf{d}$ for the entire environment is%
\begin{equation}
\mathbf{d=}\int d\mathbf{r}^{\prime }\mathbf{r}^{\prime }\Delta \overline{%
\rho }_{c}\left( \mathbf{r}\right) .  \label{5.9}
\end{equation}

These results are still quite formal but they provide the basis
for the phenomenology proposed for this part of the modeling:
Replace all effects of the classical environment on the quantum
system, exclusive of bonding, by an effective dipole representing
the polarization of the medium by the quantum domain. The origin
of this dipole in the above analysis shows that in general it will
depend on the geometry and the state of both the classical and
quantum domain (e.g., it will change under conditions of strain).
In the phenomenological application of this prescription the
dipole must be supplied by some simpler means since the electronic
contribution to $\Delta \overline{\rho }_{c}\left(
\mathbf{r}\right) $ is not known (for example, see reference
\cite{Mallik2004}).

\subsection{Electron exchange with environment}

The contributions from $\overline{\rho }_{c}\left(
\mathbf{r}\right) $ in the regions where bonds have been cut
require a more detailed treatment. Clearly, a necessary condition
is that valence saturation should be restored in the quantum
domain. However, that is not sufficient to assure the correct
charge density there nor the correct forces within that domain.
Instead, the charge density near the border ions responsible for
bonding should induce a realistic charge density within the
quantum domain. To see how this can be
done first write the contribution from one such border ion to (\ref{5.7}) as%
\begin{equation}
\int d\mathbf{r}^{\prime }\frac{1}{\left|
\mathbf{r}-\mathbf{r}^{\prime }\right| }\chi \left( \left|
\mathbf{r}^{\prime }-\mathbf{R}_{\alpha }\right| \right)
\overline{\rho }_{c}\left( \mathbf{r}\right) \equiv \phi _{p\alpha
}\left( \mathbf{r}\right)   \label{5.10}
\end{equation}%
The potential $\phi _{p\alpha }\left( \mathbf{r}\right)$
represents the actual electrostatic and exchange effects of the
nucleus of the border of the quantum domain. This is comprised
largely of the ion at the site plus its closed shell electrons
distorted by exchange and correlation effects of the site in the
quantum domain with which it is bonding. Consequently, an
appropriate pseudo-potential is introduced at each such border ion
whose behavior in the direction of the quantum system is the same
as that of the actual ion. These ions plus their pseudo-potentials
will be called ''pseudo atoms''.

To accomplish this, an appropriate candidate for the pseudo atom
is chosen based on valency of the particular pair of border ion
and its neighbor in the quantum domain. Next, a large molecule or
small cluster containing that bonding pair is chosen for training
the pseudo atom. The bond is then broken and the relevant member
of the pair replaced by the pseudo atom. The training consists of
parameterizing the pseudo atoms' effective potential to give the
same forces as in the original cluster. For example, the pseudo
atom might consist of an ion plus its closed shell electrons
chosen to satisfy the valency for the bond broken. The adjustable
parameters could refer to a characterization of the closed shell
electron distribution. In this way, it is assured that the pseudo
atom not only gives the correct saturation of the dangling bond
but also reproduces the forces \ within the cluster and hence
gives a realistic representation of the charge distribution
between the chosen pair. For the small training cluster (e.g.,
Figure \ref{fig6}), forces are determined both at equilibrium and
under strain to within about one tenth percent. The primary
assumption is that the bonding of interest is a local effect, so
that the pseudo atom trained in the cluster will have a similar
accuracy when used in the bulk solid . The training depends on the
particular quantum method used to describe the solid, and is
illustrated in more detail for the \ specific example considered
in the later sections.

In summary, the environmental effects on the quantum system are
accounted for approximately by a dipole representing its
polarization and a collection of pseudo atoms located at the sites
of ions where bonds have
been cut%
\begin{equation}
V_{eq}\rightarrow \frac{1}{2}\int_{\mathcal{Q}}d\mathbf{r}\rho
_{eq}\left(
\mathbf{r}\right) \left( \int_{\mathcal{Q}}d\mathbf{r}^{\prime }\frac{1}{%
\left| \mathbf{r}-\mathbf{r}^{\prime }\right| }\left( \rho
_{eq}\left( \mathbf{r}^{\prime }\right) -\delta \left(
\mathbf{r}-\mathbf{r}^{\prime
}\right) +\overline{\rho }_{i}\left( \mathbf{r}^{\prime }\right) \right) +%
\frac{\widehat{\mathbf{r}}\cdot \mathbf{d}}{r^{2}}+\sum_{\alpha
\subset \mathcal{V}}\phi _{p\alpha }\left( \mathbf{r}\right)
\right) .  \label{5.11}
\end{equation}%
This model now allows solution to (\ref{5.5}) for the electron
distribution the quantum domain, including its coupling to the
classical environment.
Then $\overline{\rho }_{eq}\left( \mathbf{r}\right) $ is calculated from (%
\ref{5.3}) and (\ref{5.4}). Finally, the desired potentials $V_{ic}$ and $%
V_{iq}$ of (\ref{3.6}) and (\ref{3.7}) are fully determined
\begin{equation}
V_{ic}=\frac{1}{2}\sum_{i,j}\sum_{\alpha }^{N_{i}}\sum_{\beta
}^{N_{j}}V_{ij}\left( \left| \mathbf{R}_{i\alpha
}-\mathbf{R}_{j\beta }\right| \right)
+\frac{1}{2}\int_{\mathcal{C}}d\mathbf{r}\rho _{i}\left(
\mathbf{r}\right) \int_{\mathcal{Q}}d\mathbf{r}^{\prime
}\frac{1}{\left|
\mathbf{r}-\mathbf{r}^{\prime }\right| }\left( \rho _{i}\left( \mathbf{r}%
^{\prime }\right) +\overline{\rho }_{eq}\left( \mathbf{r}^{\prime
},t\right) \right) ,  \label{5.12}
\end{equation}%
and%
\begin{equation}
V_{iq}=\frac{1}{2}\int_{\mathcal{Q}}d\mathbf{r}\rho _{i}\left( \mathbf{r}%
\right) \left( \int_{\mathcal{Q}}d\mathbf{r}^{\prime
}\frac{1}{\left|
\mathbf{r}-\mathbf{r}^{\prime }\right| }\left( \rho _{i}\left( \mathbf{r}%
^{\prime }\right) -\delta \left( \mathbf{r-r}^{\prime }\right) +\overline{%
\rho }_{eq}\left( \mathbf{r},t\right) \right) +\frac{\widehat{\mathbf{r}}%
\cdot \mathbf{d}}{r^{2}}+\sum_{\alpha \subset \mathcal{V}}\phi
_{p\alpha }\left( \mathbf{r}\right) \right) .  \label{5.13}
\end{equation}%
The solution to the classical ion motion (\ref{3.5}) can then
proceed via MD simulation. This completes the proposed scheme for
multiscale modeling of the idealized quantum solid. Rather than
critique further the assumptions made in this abstract context,
the scheme is discussed in more detail for a specific application
to an $Si0_{2}$ nanorod.

\section{$Si0_{2}$ Nanorod - A Critical Test}

The modeling scheme of the previous sections is now illustrated in
detail and tested critically for a model solid consisting of a
silica nanorod \cite {Zhu2003}, shown in Figure \ref{fig1}. It
consists of 108 Si and O ions with the stoichiometric ratio of \
one silicon to two oxygens, a stack of six $Si_{6}O_{6}$ rings
sharing both above and below a ring of oxygen atoms. To saturate
overall valency the nanorod is terminated with end cap rings
whereby each silicon atom of the terminating ring is connected by
bridging oxygen or two interstitial oxygens. The size of this
nanorod can be readily adjusted by adding or removing
$Si_{6}O_{6}$ planar rings with corresponding O atoms. Many of the
results reported here have been studied for larger rods as well
\cite {Zhu2003}. As noted in the Introduction, this system
exhibits strain to fracture (Figure 1), yet it is small enough for
practical application of the chosen quantum mechanics to the whole
rod. Consequently, the results of modeling it as a composite rod
can be tested quantitatively against the ''exact'' results. That
is the objective of this section.

The decomposition of the rod into classical and quantum domains is
accomplished by identifying the ions of one of the rings near the
center as the QM domain. The rest of the ions define the classical
domain. The actual geometry of the interface is chosen by charge
neutrality of the quantum domain, as given by Eq.(\ref{3.4}).

\subsection{Transfer Hamiltonian Quantum Mechanics}

At the quantum level, each possible choice for the quantum
chemical method entails different approximations used to optimize
the accuracy and speed of the calculations. Each approximation has
its own advantages and limitations, but a general embedding scheme
should be insensitive to these. Much of the previous work on
multiscale modeling is based on tight binding models
\cite{Broughton1999,Rudd2000} because of their ability to treat
hundreds of atoms in the QM domain. However, the Hamiltonian in
these models is oversimplified and cannot account for charge
transfer, which is essential to bond breaking. A more complete
quantum description is considered here. It is appropriate to
stress at this point that the objective of the following sections
is to demonstrate and test the modeling scheme for a given quantum
description. That is a separate issue from the question of how
accurate that description may be. To test how robust the proposed
embedding scheme is, two different methods are chosen for the
underlying quantum mechanical description. The first method is the
Transfer Hamiltonian (TH) - Neglect of Diatomic Differential
Overlap (NDDO) \cite{Taylor2003}. The second method is the
Born-Oppenheimer local spin density \cite{Barnett1993} within
density functional theory using a generalized gradient
approximation. The results obtained from the TH are described
first.

The goal of the TH strategy is to provide forces for realistic MD
simulations of the quality of the coupled cluster singles and doubles method %
\cite{Bartlett1995} in computationally accessible times
\cite{Taylor2004}. The TH is a low rank, self - consistent,
quantum mechanical single particle operator whose matrix elements
are given in terms of parameterizable functions. The functional
form of the TH can be chosen to be of any semi-empirical
functional form, but here the NDDO form is chosen because of its
better qualitative description over Intermediate Neglect of
Differential Overlap (INDO) Hamiltonians \cite{Hsiao2001}. The
NDDO approximation restricts charge distribution to basis
functions on the same center, so this has at most two-atom
interactions. Because of its NDDO form it is anticipated that the
parameters fit to coupled cluster data are saturated for small
cluster size and that these parameters can then be transferred to
extended systems. This transfer might be considered an example of
serial multiscale-modeling.

To apply the TH to silica systems for studying complex processes
such as fracture or hydrolysis where the role of electrons is
important, Taylor \textit{et al.} \cite{Taylor2003} parameterized the TH for the following reactions $%
Si_{2}H_{6}\rightarrow 2SiH_{3}$, $Si_{2}O_{7}H_{6}\rightarrow
Si(OH)_{3}+SiO(OH)_{3}$ (radical mode of dissociation), and $%
Si_{2}O_{7}H_{6}\rightarrow $ $^{+}Si(OH)_{3}+^{-}OSi(OH)_{3}$
(ionic mode of dissociation). By describing the two different
dissociation pathways for $Si_{2}O_{7}H_{6}$, electron state
specificity is systematically introduced into the TH model. Since
fracture is expected to involve a variety of dissociations this TH
model is expected to be appropriate for study of strained
conditions as is done here. Thus the parameterized TH model is
expected to provide accurate results for such processes, while
requiring only the computational intensity of semi-empirical
methods which are orders of magnitude less demanding than coupled
cluster calculations on the same systems. In the next few
subsections, the charge densities and forces that arise from the
TH are the reference data by which the proposed method for
multiscale modeling is judged.

\subsection{Pair Potentials for the Classical Domain}

The method that we propose for constructing a classical pair
potential has been previously presented in detail
\cite{Mallik20042}, so only a brief summary is given here. The
functional form for the potential is chosen to be one that is
commonly used in the literature. In the case of silica, this is
the TTAM \cite{Tsuneyuki1998} and BKS \cite{Beest1990} forms which
are pair potentials comprised of a Coulombic interaction, a
short-range exponential repulsion, and a van der Waals attraction
(the last two terms are collectively known as the 'Buckingham'
potential)
\begin{equation}
V_{ij}(r)=\frac{q_{i}q_{j}}{r}+a_{ij}e^{-b_{ij}r}-\frac{c_{ij}}{r^{6}}
\label{6.1}
\end{equation}%
There are 10 free parameters as the charge neutrality of the
$SiO_{2}$ unit requires that $q_{Si}= -2q_{O}$, so that $a_{ij}$,
$b_{ij}$, and $c_{ij}$ for each atom pair and one charge define
the potential. For the purposes here, this potential is used to
generate the dynamics (MD simulations) so that forces have been
used as the property "training" the potential (determining the
free parameters). More specifically, force data from the silica
nanorod calculated with the TH at equilibrium and a few small
strain configurations have been used.

\begin{figure}[ht]
\includegraphics[width=9cm]{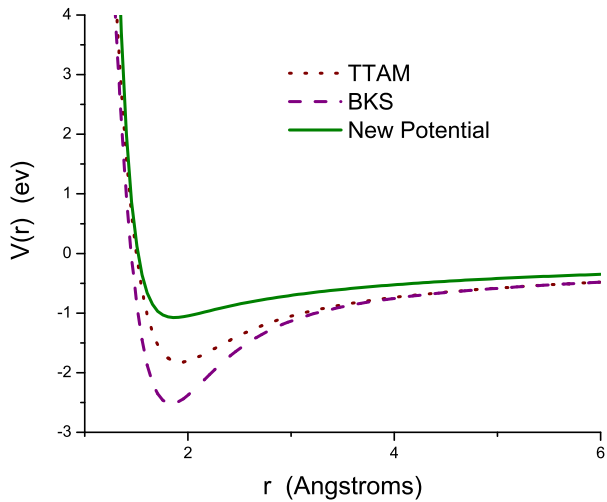}
\caption{(color online) Si-O pairwise interactions for the three potentials}
\label{fig2}
\end{figure}

Fitting a set of ten parameters from hundreds of data points
requires an optimization approach that is capable of exploring a
large parameter space efficiently without being trapped in local
minima. For this task, a genetic algorithm (GA) has been chosen as
detailed in reference \cite{Mallik20042}. The GA generates a
number of sets of parameters of a given fitness which are then
tested using a standard geometry optimization technique (BFGS \cite%
{Zhu1994}) to ensure that the parameterization gives a stable
equilibrium configuration. The parameter set that gives a geometry
closest to the TH nanorod is then chosen and rescaled to reproduce
the equilibrium structure to within a few percent. (See the
appendix of reference \cite{Mallik20042} for details of the
rescaling used.) Table \ref{tab1} presents the final parameters
obtained compared to the TTAM and BKS parameters. Table \ref{tab2}
shows that the new classical potential reproduces the nanorod
structure to within a few percent compared to larger discrepancies
with the two standard potentials. The shapes of these three
potentials are shown in Figure \ref{fig2} for the Si-O pair
interaction.

\begin{figure}[ht]
\includegraphics[width=9cm]{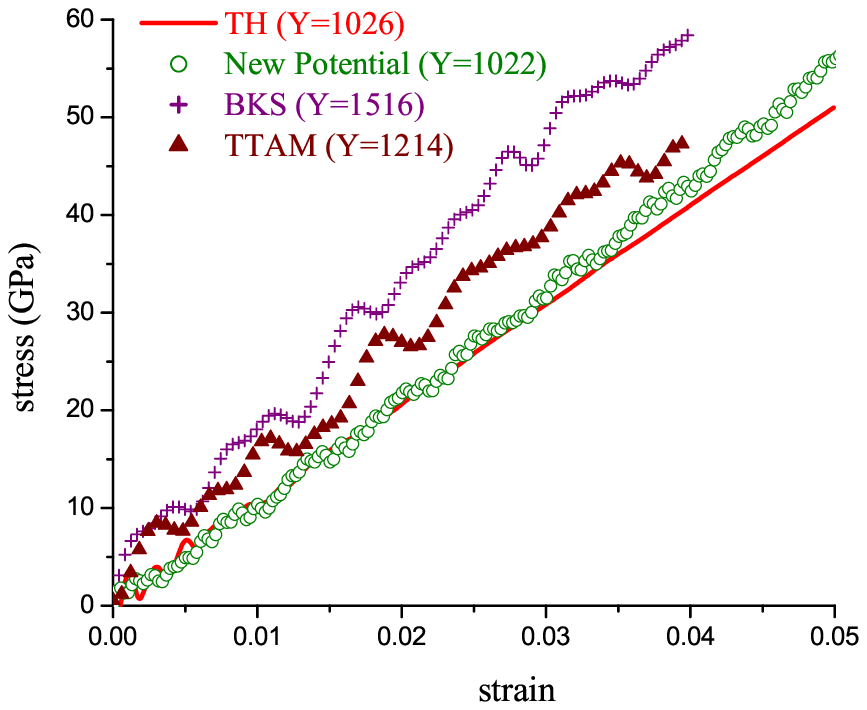}
\caption{(color online)Stress-strain curves for the different potentials and for
TH quantum mechanics} \label{fig3}
\end{figure}

Finally, the small strain elastic behavior of this new potential
is studied. This is accomplished using MD simulation of uniaxial
strain along the long axis (z) of the silica nanorod at a strain
rate of 25 m/s. The integration is done using 2 fs time steps and
a temperature of 10 K enforced by velocity rescaling. Figure
\ref{fig3} shows the resulting stress-strain curves for small
strains up to $5\%$. It is noted that the new potential indeed
reproduces the linear elastic response (Young's modulus) behavior
of the underlying TH while the TTAM and BKS potentials are
somewhat stiffer.

A key to successful multiscale modeling lies in the consistent
embedding of a QM domain in its classical MD region. This requires
that the CM region have the same structure and elastic properties
as the QM domain. The potential obtained in this section meets
these criteria for the silica nanorod. Essentially, a classical
representation of the TH silica nanorod has been found for both
equilibrium and near-equilibrium configurations.

\begin{table}
\caption{\label{tab1}Potential Parameters.}
\begin{ruledtabular}
\begin{tabular}{cccc}
Parameter&TTAM&BKS&New potential  \\
\hline
$q_{si}$& 2.4& 2.4 &2.05 \\
$q_{o}$ & $-1.2$ & $-1.2$ & $-1.025$\\
$a_{oo}$& $1756.98$ & $1388.773$ &39439.87 \\
$b_{oo}$& $2.846$ & $2.760$ & $4.66$ \\
$c_{oo}$ & $214.75$ & $175.00$ & $5.85$ \\
$a_{osi}$& $10722.23$ & $18003.757$ & $240101.9$ \\
$b_{osi}$ & $4.796$ & $4.873$ & $7.88$\\
$c_{osi}$ & $70.739$ & $133.538$ & $2.06$ \\
$a_{sisi}$& $8.73E+8$ & $0$ & $27530.45$ \\
$b_{sisi}$ & $15.22$ & $0$ & $21.42$ \\
$c_{sisi}$ & $23.265$& $0$ & $1.55$ \\
\end{tabular}
\end{ruledtabular}
\end{table}

\begin{table}
\caption{\label{tab2}Structure of the Nanorod from the Different
Potentials.}
\begin{ruledtabular}
\begin{tabular}{cccccccc}
Bond lengths and angles ($<$)& TH & New TH potential & $\%$ error & TTAM & $\%$ error & BKS & $\%$ error  \\
\hline
 & & & \textit{In Silica Plane} & & & & \\
 $Si-O$ & $1.641$ & $1.642$ & $0.04$ & $1.65$ & $0.76$ & $1.611$ & $1.5$ \\
 $< Si-O-Si$ & $170.06$ & $173.7$ & $2.14$ & $162.9$ & $4.2$ &
 $161.1$& $5.2$ \\
 \hline
 & & & \textit{Between Planes} & & & & \\
 $< Si-O-Si$ & $103.8$ & $104.3$ & $0.4$ & $104.6$ & $0.7$ &
 $104.9$& $1.04$ \\
 \hline
 & & & \textit{End caps} & & & & \\
$Si-O$ & $1.71$ & $1.67$ & $2.5$ & $1.67$ & $2.5$ & $1.62$ & $5.1$ \\
 $< Si-O-Si$ & $102.03$ & $103.2$ & $1.14$ & $100.6$ & $1.45$ &
 $100.8$& $1.23$ \\
 \hline
 Length & $16.49$& $16.34$& $0.9$& $16.31$& $1.1$& $15.86$& $3.8$
 \\
 Diameter & $6.55$& $6.54$& $0.15$& $6.57$& $0.35$& $6.41$& $2.0$
 \\
\end{tabular}
\end{ruledtabular}
\end{table}

\subsection{The Quantum Domain}

A proper description of the quantum domain requires incorporating
two kinds of environmental effects: a short-range electronic
exchange interaction and long-range Coulombic interactions. Most
previous studies of QM/CM simulations have focused on the former
and have neglected the long-range interactions. It will be shown
here that both kinds of interactions must be taken into account
for an accurate description of forces and charge densities in the
QM domain. This subsection first reviews briefly the various
termination schemes used in the literature to treat the bond
cutting region. Then, the method proposed here for construction of
a pseudo-atom to saturate dangling bonds is described, together
with modeling the rest of the CM environment by lowest order
multipoles. The charge densities and forces in the QM domain
obtained from our scheme of pseudo-atoms and dipoles with those
obtained from the conventionally used link atoms (bond saturation
with hydrogen atoms) and dipoles are compared to the "exact" TH
reference data for the entire nanorod.

Over the past decade there
have been numerous proposals for different types of termination
schemes to accommodate dangling bonds. Only a few of the more
relevant ones are noted for context.
(i) Link Atom method: This commonly used method is the Link Atom
method, presented by Singh and Kollman \cite{Singh1986}. In this method, hydrogen
atoms are added to the CM side of broken covalent bond to satisfy
the valency of the QM system. There are many variations within the
implementation of the LA method, for example, the double Link Atom
method \cite{Das2002}, the Add-Remove Link Atom method \cite{Swart2003} or the
Scaled-Position-Link-Atom Method (SPLAM)\cite{Eichinger1999}. In the present work,
the hydrogen atoms are placed at a fixed distance of 0.97 \AA\ from the
cut Si bond.
(ii) Connection Atom method \cite{Antes1999}: A connection atom is developed to saturate a C-C bond,
such that the connection atom mimics the effect of a methyl group. This connection
atom interacts with the QM atom quantum mechanically and the interactions with the
CM atoms are handled classically using a carbon force field. The parameters of the
connection atom are determined using semi-empirical methods \cite{Stewart1990} such as AM1, MNDO or PM3
designed to reproduce theoretical QM data for energies, geometry and net charges.
About 30 different methyl hydrocarbons were used as reference molecules. The parameters
adjusted are the orbital exponent , one-center one-electron energy , one center
two-electron integral , resonance parameter , and repulsion term . The mathematical
functional forms of these parameters can be found in any book on semi-empirical theory \cite{Stewart1990}
or papers by Dewar and Theil \cite{Dewar1976}.

There are numerous other termination schemes for the QM/CM
boundary like the 'pseudobond' scheme \cite{Zhang1999}, `IMOMM'
\cite{Maseras1995}, \cite{Humbel1996} and
'ONIOM'\cite{Svensson1998}, \cite{Maseras1995},
\cite{Dapprish1999} procedures , 'effective group potential (EGP)'
\cite{Poteau2001}. However these methods are not discussed here.

\begin{figure}[ht]
\includegraphics[width=8cm]{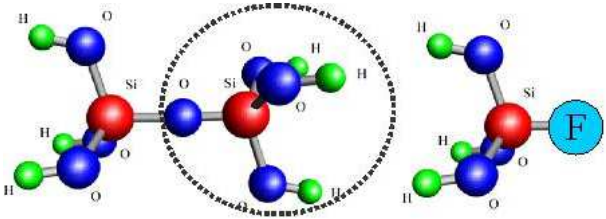}
\caption{(color online)Training of Pseudo-atom on Pyrosilicic acid}
\label{fig4}
\end{figure}

The method proposed here will be referred to as the
''pseudo-atom'' method. In the present case it is based on the TH
approach for saturating a bond terminating in Si (in silica
systems). The training cluster is a pyrosilicic acid molecule
(Figure \ref{fig4}). The part of the molecule within the dotted
lines is replaced by a fluorine (F) atom whose NDDO parameters are
then adjusted to give the correct QM forces (which implies correct
geometry as well) and charge density in other parts of the
molecule (outside the dotted lines). The pseudo atom is placed at
the same position as the neighboring CM atom (O in this case) in
the bond being cut and hence geometry optimization as in the LA
method is not required. The specific NDDO parameters modified are:
one-center-one-electron integrals ,  Coulomb integrals,  exchange
integral, and two-center-one-electron resonance integrals
\cite{Dewar1976}. This method is similar to the previously described
connection atom developed in reference \cite{Antes1999}. However
the advantage of using pseudo-atoms is that they are trained to
give the correct QM forces and charge densities for both the
equilibrium Si-O bond being cut and for small distortions (up to
$3-4\%$ from the equilibrium) in the Si-O bond length as well.
This allows use of the pseudo-atoms even while studying dynamics
in the system.
To test this method, the actual charge density from the TH method
is calculated instead of the more commonly used Mulliken
populations \cite{Mulliken1955}, which are known to have several
common problems ( e.g., equal apportioning of electrons between
pairs of atoms, even if their electronegetavities are very
different).

\begin{figure}[ht]
\includegraphics[width=6cm]{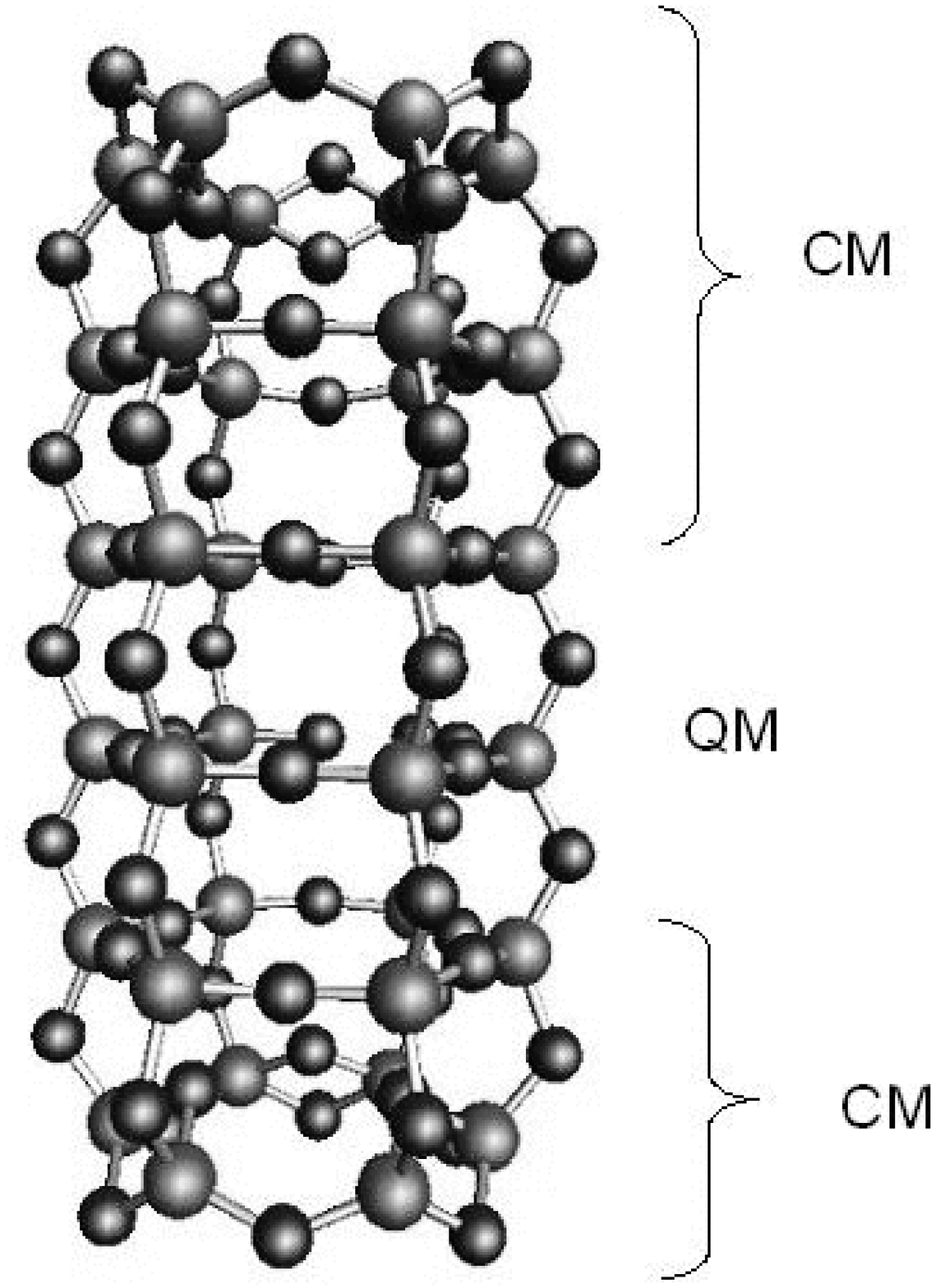}
\includegraphics[width=8cm]{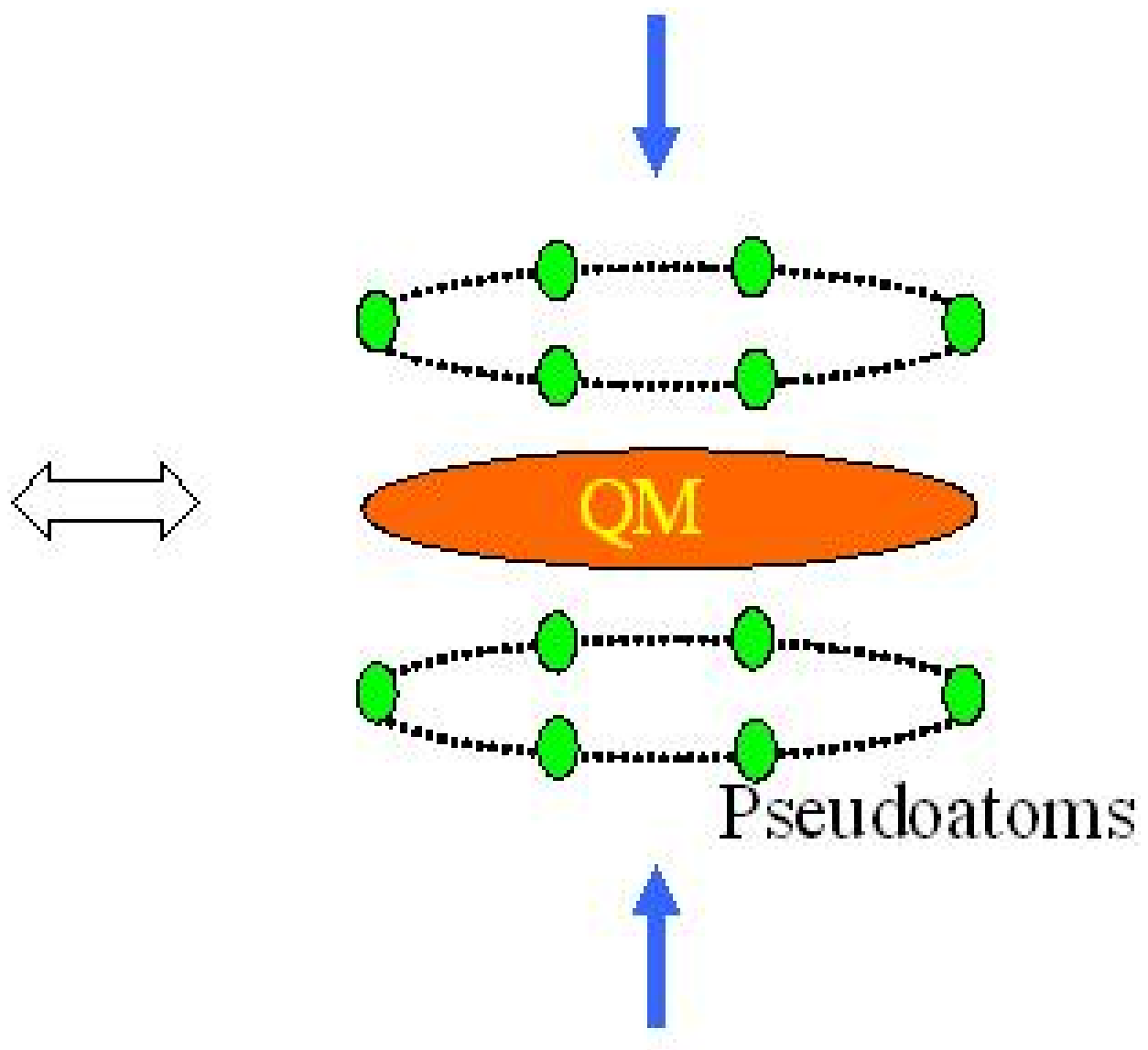}
\caption{(color online) Embedding Scheme: Approximating the CM region by
pseudo-atoms and dipoles} \label{fig5}
\end{figure}

\begin{figure}[ht]
\includegraphics[width=9cm]{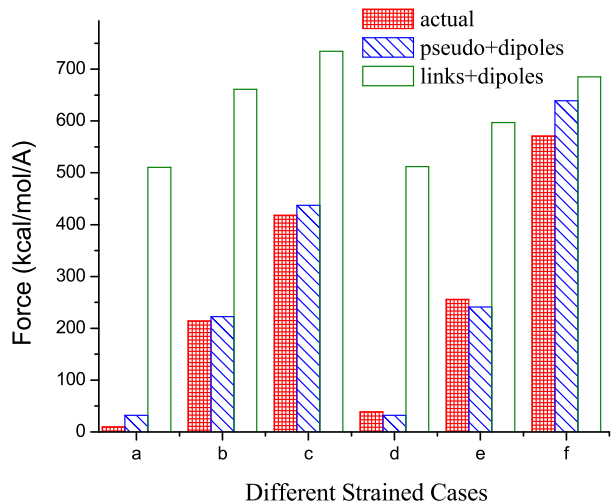}
\caption{Forces on Si nuclei for various cases studied}
\label{fig6}
\end{figure}

The remainder of the environment is represented as two dipoles for
the top and bottom portions of the rod (Figure \ref{fig5}). The
values of the dipole have been calculated using the TH-NDDO charge
density for these two portions of the rod. These domains are taken
to be charge neutral, but are polarized by the presence of the QM
domain. The validity of the approximation by dipoles has been
checked by comparing the force on an Si nuclei of the ring due to
all charges and that due to the dipole,  with excellent agreement.
Constructing the full system as a composite of a QM domain
terminated by pseudo atoms and embedded in dipolar fields leads to
the forces displayed in Figure \ref{fig6}. It is found
that the normalized difference of charge
density is reproduced to within $0.1 \%$
in the plane of the ring \cite{Mallik2004}.

This scheme was found to be applicable to both equilibrium and
strained configurations, as illustrated in Figure \ref{fig6} for
the following cases: a) equilibrium, b) the ring of the QM domain
radially expanded by $5\%$, and c) a distorted ring in which one
Si atom is radially pushed out and one Si pushed in. Also shown
are the corresponding results for a longer 10 ring rod (cases
d),e), and f)).

The pseudo-atoms and the dipoles reproduce the force in the QM
domain within $1\%$ error. In contrast, the LA placed at 0.97 \AA\
from the terminating Si atom and along the Si-O bond lead to a
large force on the Si atom at equilibrium, and generally poor
results for the strained cases. One might argue that the LA method
could be improved if placed at ''optimal'' positions. To test
this, both the Si-H bond length as well as the alignment of the LA
was varied to give minimum force on the Si atom. This was obtained
when the LA is placed at a distance of 1.45 \AA\ from the silicon
and the bond angle is decreased by about 5 degrees. Although the
LA at this position gives forces comparable to those for
pseudo-atoms plus the dipoles, it fails to reproduce the correct
charge densities.

\subsection{The Composite Rod}

The composite rod is built by embedding the QM region in its CM
environment as described in Section III. The forces on the atoms
in the CM region are calculated from the pair potential developed
in IV B above, while the forces on the atoms in the QM domain are
obtained  from the charge density calculated by the TH method as
described in VI C . An important test of the composite rod is its
indistinguishability from the TH rod for near equilibrium states,
i.e. its structure and elastic properties.

\begin{figure}[ht]
\includegraphics[width=9cm]{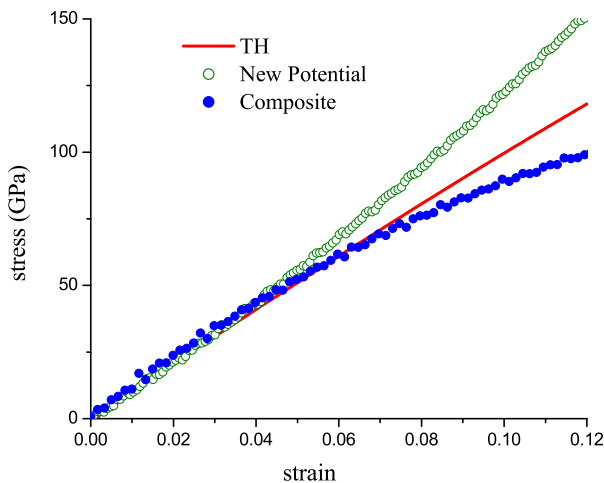}
\caption{(color online) Stress-strain curve for the TH (quantum), new potential
(classical), and composite rods} \label{fig7}
\end{figure}

The structural properties (bond lengths, bond angles) have an
accuracy comparable to that of the purely classical rod described
using this pair potential (Table \ref{tab2}). Consequently,
attention here will be focused on the elastic properties. Figure
\ref{fig7} shows the stress-strain behavior of the nanorod
obtained from three different methods: (i) quantum mechanics TH
method for the entire rod, (ii) pair potentials for the entire
rod, and (iii) the composite rod constructed as described above.
The three overlaying curves (measured at 0.01K) indicate that the
composite rod is identical to the rod obtained from TH and the
pair potential nanorod in terms of small strain elastic properties
and structure. The stress-strain results shows the success of our
multiscale method indicating that the composite rod is
indistinguishable from the underlying quantum mechanics for states
near equilibrium.

\subsection{Notched Nanorod }

The elastic properties of the composite rod obtained from both of
these potentials do not agree beyond $10\%$ strain. This is
because the pseudo-atoms, trained at regions only close to the
equilibrium configuration, fail to give the correct charge
densities at such high strains. This diagnosis was checked by
comparing the charge density in the QM domain of the $12\%$
strained rod with that of equilibrium configuration and a
difference of $6\%$ was found. Retraining of the pseudo-atoms
improve the stress-strain performance of the composite rod beyond
a strain of $10\%$ is possible, but this was not done because of
the large strains involved. Real systems like glasses have many
inherent defects which act as stress concentrators that cause the
material to break at much lower strains than observed for the
nanorod (with a yield stress of about 190 GPa).

\begin{figure}[ht]
\includegraphics[width=6cm]{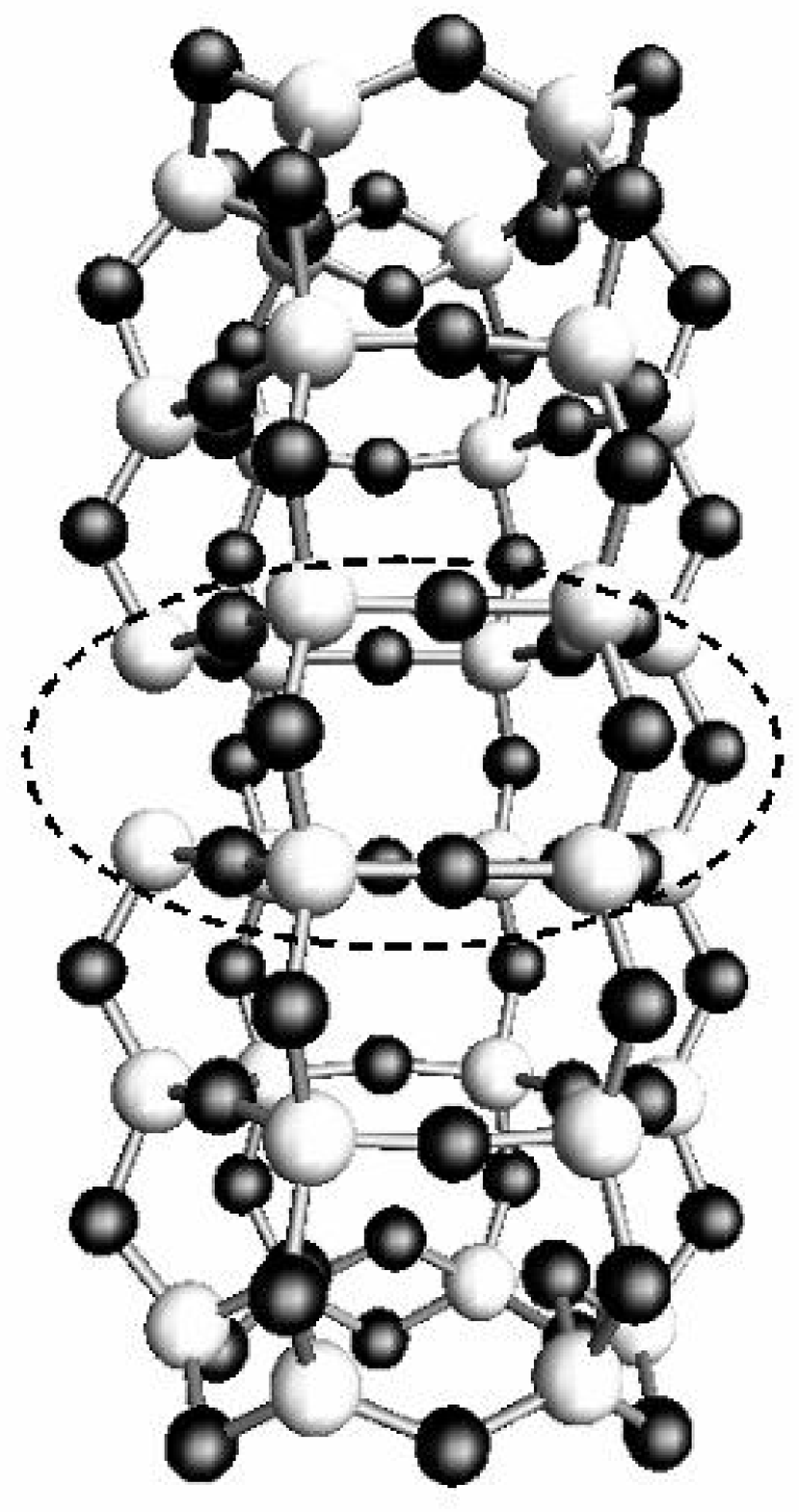}\hspace{1.0 cm}
\includegraphics[width=8cm]{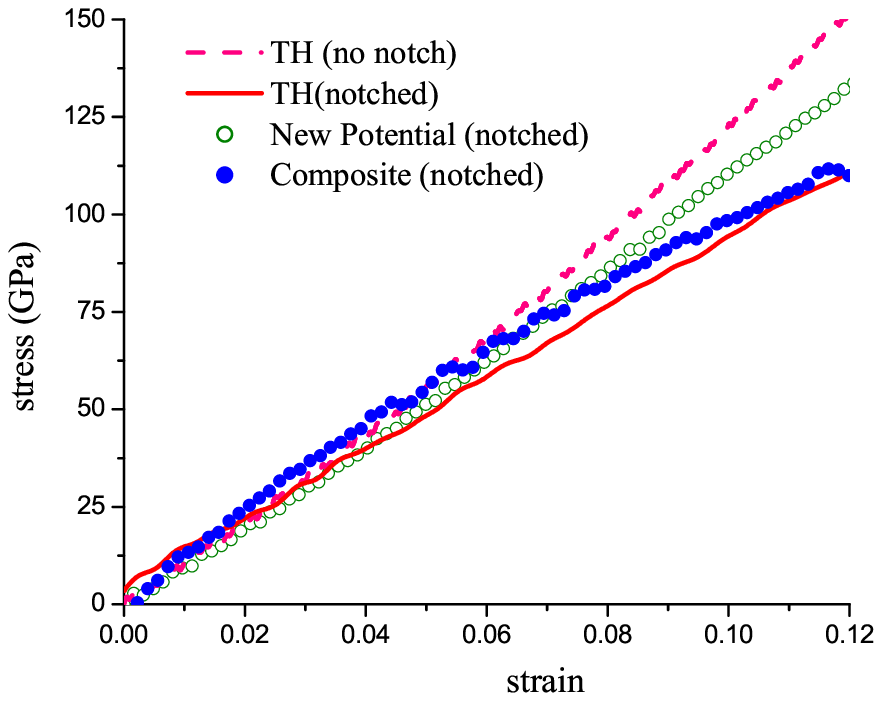}
\caption{(color online) The notched nanorod and corresponding stress-strain
curve} \label{fig8}
\end{figure}

To illustrate this, a defect notch was placed in the 108 atom nanorod by removal of an
oxygen atom as shown in Figure \ref{fig8}. The MD stress-strain
curves for this notched rod were found using TH quantum mechanics,
the trained classical potential, and the composite. We note that
the presence of only a small defect can significantly reduce the
yield stress of the material and make it more prone to fracture.
The TH curve for the defect-free rod is plotted in the same figure
to contrast the value of the yield stress. As can be seen there is
a reduction of \symbol{126}60 GPa in the yield stress.
For the composite rod in this case, the QM domain was chosen to
consist of 2 silica planes and the intermediate 5 oxygen atoms
(see Figure \ref{fig8}), so that the defect could be located in
the QM region. The stress - strain curve for this composite
notched rod agrees well with that for the TH quantum calculation
up to 10\% strain. Above 4\% strain the curve for the composite
notched nanorod follows that of the TH instead that of the trained
classical potential nanorod, showing that the composite rod is
representing the ''real'' material. This is exactly what is
required of multiscale modeling.

\subsection{DFT Quantum Mechanics}

A proper multiscale modeling procedure should be independent of
the choice of underlying quantum mechanical method. To test the
method proposed here, the analysis of subsections (B)-(D) is
repeated using a density functional theory (DFT) instead of the
Transfer Hamiltonian as the quantum mechanical approximation. The
primary results of this section are a confirmation that the
isolation of the QM domain with pseudo-atoms and dipoles, plus the
construction of a classical potential based on the DFT forces
leads to accuracies of the same quality as those described already
using the TH quantum mechanics. Hence, the multiscale modeling
scheme is faithful to the chosen form for the underlying quantum
mechanics in both cases.
    The DFT code used is a parallel multiscale program
package, known as Born-Oppenheimer molecular dynamics ``BOMD''
\cite{Barnett1993}. It is a generalized gradient approximation
(GGA) within local spin DFT (LSDFT). A Troullier-Martin
Pseudo-Potential \cite{Troullier1991a,Troullier1991b} is used for
the effect of the chemically inert core states on the valence
states. The code employs the dual space formalism for calculation
of the DFT energy. A plane wave basis set and cut-off energy of
30.84 Rydbergs is chosen.

\begin{figure}[ht]
\includegraphics[width=9cm]{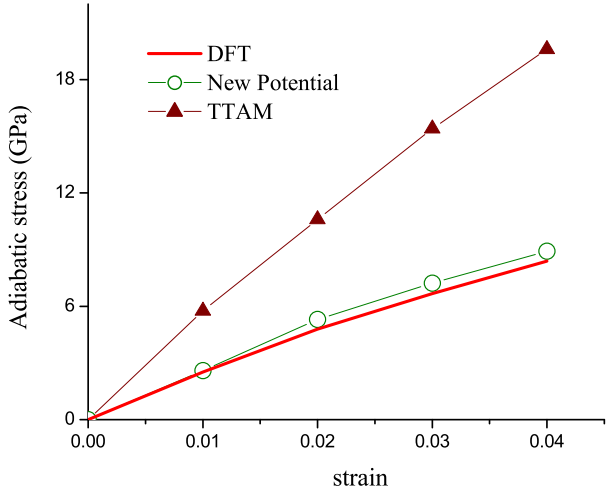}
\caption{(color online)Adiabatic stress curves for DFT, DFT-potential, and TTAM
potential} \label{fig9}
\end{figure}

A pseudo-atom for the quantum domain is constructed based on
parameterization of the Troullier-Martin (TM) pseudopotential,
using a cut-off radius of $1.5 \AA$. Once again the pyrosilicic
acid molecule is chosen for parameterization of the fluorine-like
pseudo atom and its position is constrained to be at the same
place as the O atom (Figure \ref{fig4}). Unlike the TH-NDDO method
in which both the electron-ion and electron-electron interaction
parameters could be changed , in DFT only the electron-ion
interactions can be modified. The three options to alter these
interactions for the F atom using the TM pseudopotential are: (i)
the core charge on F, (ii) omission of the non-local part in the
potential, and (iii) switching the local and non-local part
between s and p orbitals. All three possible choices and their
combinations were explored to find the optimal reproduction of
forces on the terminated Si atom in the pyrosilicic acid. The best
results were obtained when the core charge is 7.0 and the
non-local part is omitted. This pseudo-atom was then applied to
the composite nanorod constructed as above (Figure \ref{fig5}). The values
of the dipoles were recalculated from the charge density obtained
using the DFT results for the CM portions of the rod. The force on
a Si atom in the QM domain was calculated for the rod in
equilibrium and all the strained cases considered in subsection C.
The results obtained from a DFT calculation on the whole rod are
now taken to be the "exact" reference forces. It is found that
the forces and the charge densities in the QM domain
can be generated using DFT pseudo-atoms and dipoles to $1\%$
accuracy.

    Next, a new potential having the same form as TTAM is constructed
to predict the same structure and elastic properties as for the QM
rod. A GA with DFT force data up to $4\%$ expansion followed by
the scaling procedure is used as described in section VI B above to
find the parameters for the potential. The charge on the ions is
lower (as was found for TH quantum mechanics) than that given in
the TTAM and BKS potentials. Also, the van der Waals interaction
is much weaker than in those potentials. This is expected since
the DFT forces fail to represent this effect, falling
exponentially rather than algebraically with separation. The
parameters for the DFT potential and a comparison of the resulting
nanorod structure with that for the BKS and TTAM nanorods are
given in reference \cite{Mallik20042}. The agreement between the
results from the DFT potential and DFT quantum mechanics is
similar to that found in Table \ref{tab2} for the TH quantum
mechanics.

A stress strain curve for the entire rod using MD with DFT quantum
mechanics is computationally too intensive. Instead, only selected
equilibrium and adiabatic strain configurations were calculated
with DFT. The equilibrium structure was determined by sequential
DFT calculations and nuclear relaxation to find the minimum energy
configuration. The strained configurations were obtained from an
affine transformation of the minimum energy configuration by 1, 2,
3, and 4 \%, with a single DFT calculation of forces at each of
the expanded configurations. Then the average force on the Si atom
in each ring of the DFT nanorod was computed for these four cases.
The stresses for adiabatic configurations using DFT for the entire
rod, the constructed DFT potential, and the TTAM potential were
then compared (the values of stress obtained using BKS potential
are similar to those of TTAM potential). The results are shown in
Figure \ref{fig9}.

These results suggest that the multiscale modeling method proposed
here is accurate for a wide range of possible choices for the
underlying quantum mechanical method employed.

\section{Summary and Discussion}

The objectives here have been three-fold. The first was to
describe the formal quantum structure for a real solid from which
a practical model should be obtained through a sequence of
well-identified (if not fully controlled) approximations. The
second was to propose a method for partitioning this structure
into classical and quantum domains while preserving the properties
of interest for the quantum structure in the replica composite
solid. The final objective was a quantitative test of the proposed
method by its application to a non trivial mesoscopic ''solid'',
the $SiO_{2}$ nanorod.

For the first objective, coupled Liouville - von Neumann equations
for the reduced density operators for the ions and electrons of
the system were considered. Each is coupled to the other through
the mean charge density of the complementary subsystem.
Determination of these charge densities then becomes the central
problem for further analysis. For the heavy ion component analysis
by classical MD simulation provides a practical and accurate
approach. The necessity for a detailed quantum treatment of the
electronic charge density is the primary bottleneck for progress.
For bulk samples of interest, direct application accurate quantum
chemical methods are precluded, so composite constructs with
smaller quantum subdomains are a potentially fruitful compromise.
Here, the isolation of a quantum subdomain was accomplished by
defining the reduced density matrix for electrons in the vicinity
of a selected small set of ions. The corresponding Liouville - von
Neumann equation describes the dynamics of those electrons and
ions coupled to the remainder of the solid (the classical domain).
Its environment is entirely characterized by its mean charge
density. This formulation and emphasis on the charge density is a
guiding feature of the subsequent approximations and modeling.

The specific method for constructing the composite solid consists
of modeling the charge densities of the environment for the two
subsystems. For the environment of the quantum domain the charge
density is separated into that border component responsible for
electronic bonding across the border, and a longer range Coulomb
interaction. The former is treated in detail by pseudo atoms
trained on smaller clusters to provide highly accurate local
forces and charge densities. The latter is represented by leading
order terms in a multi-pole expansion. The interactions within the
classical domain are described phenomenologically by pair
potentials. These pair potentials are designed specifically to
represent selected properties determined by the quantum chemical
method assumed to represent the solid. This presumes conditions
such that the chosen properties have such a classical
representation (e.g., near equilibrium states), and the method
constitutes a ''tuning'' of the potential to match the quantum
description on the other side of the border.

There are several critical tests of this approach, all requiring a
benchmark system for which the ''exact'' quantum data for the
entire solid is required. This system is provided here by the
nanorod for which the global quantum calculations are possible.
Then, choosing a central ring as the quantum subdoman and the
remainder as the classical domain many quantiative tests are
possible. Some of these provided here are:

\begin{enumerate}
\item The charge density and forces in the central ring computed from the
model are accurate to within one percent, both at equilibrium and
strains up to 5\%.

\item The pair potential fit to quantum data was used to construct the
entire nanorod. The resulting structure and elastic properties up
to 5\% strain were indistinguishable from those of the quantum
nanorod.

\item The proposed method was used to construct a composite classical /
quantum nanorod. Again the structure and elastic properties were
indistinguishable from those of the quantum nanorod.
\end{enumerate}
These and other results presented above constitute a demonstration
that the multiscale modeling is not creating a new solid, but
rather is faithful to the real system of interest. Similar (and
perhaps more physical) results were obtained for the nanorod with
a defect (missing Oxygen). Finally, the entire multiscale analysis
was repeated using a quite different choice for the underlying
quantum mechanics with the same degree of accuracy.

The predictions of multiscale modeling for conditions of interest,
states far from equilibrium, rely on the assumption that the
properties calculated in the quantum subdomain are indeed those of
the given quantum structure. This means that the quantum imbedding
and representation of the classical domain are ''passive'', i.e.
no new physical effects have been introduced. The method described
here manifestly satisfies this constraint for the benchmark
nanorod, and provides some concrete support for its use in
realistic applications.

\section{Acknowledgements}
This research was supported by NSF-ITR under Grant No.
DMR-0325553. The authors are indebted to S. B. Trickey, K.
Muralidharan, and D. E. Taylor for helpful discussions and
assistance.

\end{document}